\documentclass[12pt]{article}

\usepackage{epsf}
\usepackage{amsmath}
\usepackage{cite}
\usepackage{graphics}
\usepackage{amsfonts}
\usepackage{amssymb}
\usepackage{latexsym}
\usepackage{color}
\input{colordvi.tex}

\renewcommand{\thefootnote}{\#\arabic{footnote}}

\begin{document}
\setcounter{footnote}{0}

\begin{titlepage}

\begin{center}

\hfill July 2008\\

\vskip .5in

{\Large \bf
Primordial Curvature Fluctuation and Its Non-Gaussianity
in Models with Modulated Reheating  

}

\vskip .45in

{\large
Kazuhide Ichikawa$\,^1$,
Teruaki Suyama$\,^2$,
Tomo Takahashi$\,^3$ \\
and Masahide Yamaguchi$\,^{4,5}$
}

\vskip .45in

{\em
$^1$
Department of Physics and Astronomy, \\
University College London, London WC1E 6BT, UK\\
$^2$
Institute for Cosmic Ray Research, 
University of Tokyo, Kashiwa 277-8582, Japan\\
$^3$
Department of Physics, Saga University, Saga 840-8502, Japan \\
$^4$
Department of Physics and Mathematics, \\
Aoyama Gakuin University, Sagamihara 229-8558, Japan \\
$^5$
Department of Physics, Stanford University, Stanford CA 94305 \\ 
}

\end{center}

\vskip .4in

\begin{abstract}
  We investigate non-Gaussianity in the modulated reheating scenario
  where fluctuations of the decay rate of the inflaton generate
  adiabatic perturbations, paying particular attention to the
  non-linearity parameters $f_{\rm NL}, \tau_{\rm NL}$ and $g_{\rm NL}$ as well as the scalar spectral index and tensor-to-scalar
  ratio which characterize the nature of the primordial power
  spectrum.  We also take into account the pre-existing adiabatic
  perturbations produced from the inflaton fluctuations.  It has been
  known that the non-linearity between the curvature perturbations and
  the fluctuations of the decay rate can yield non-Gaussianity at the
  level of $f_{\rm NL} \sim \mathcal{O}(1)$, but we find that the
  non-linearity between the decay rate and the modulus field which
  determines the decay rate can generate much greater non-Gaussianity.
  We also discuss a consistency relation among non-linearity
  parameters which holds in the scenario and find that the modulated
  reheating yields a different one from that of the curvaton model.
  In particular, they both can yield a large positive $f_{\rm NL}$ but
  with a different sign of $g_{\rm NL}$.  This provides a possibility
  to discriminate these two competitive models by looking at the sign
  of $g_{\rm NL}$.  Furthermore, we work on some concrete inflation
  models and investigate in what cases models predict the spectral
  index and the tensor-to-scalar ratio allowed by the current data
  while generating large non-Gaussianity, which may have many
  implications for model-buildings of the inflationary universe.

\end{abstract}
\end{titlepage}

\renewcommand{\thepage}{\arabic{page}}
\setcounter{page}{1}
\renewcommand{\thefootnote}{\#\arabic{footnote}}

%%%%%%%%%%%%%%%%%%%%%%%%%%%%%%%%%%%%%%%%%%%%%%%%%%%%%%%%%%%%%%%%%%%%%%%%%%%
\section{Introduction}
%%%%%%%%%%%%%%%%%%%%%%%%%%%%%%%%%%%%%%%%%%%%%%%%%%%%%%%%%%%%%%%%%%%%%%%%%%%

Inflation is a promising candidate to generate primordial density
fluctuations as well as to solve the horizon and the flatness
problems. Although many observational supports for the inflationary
scenario have been accumulated, it is still unclear what mechanism is
really responsible for producing primordial density fluctuations.  We
usually assume they are generated by fluctuations of the inflaton, but
those of other scalar fields can produce the primordial fluctuations
too. Such scalar fields generically exist in the extensions of the
standard model of particle physics, which motivates the curvaton
scenario \cite{Enqvist:2001zp,Lyth:2001nq,Moroi:2001ct} and modulated
reheating scenarios \cite{Dvali:2003em,Kofman:2003nx}.  One of
interesting features of these scenarios is that primordial
non-Gaussianity can possibly be very large, in contrast to the case of
single field inflation models where only the inflaton is responsible
for density fluctuations so that almost perfect Gaussian fluctuations
arise.  Therefore, non-Gaussianity can be a very powerful tool to
identify the real source of the primordial fluctuations.  In fact,
there has been reported that non-Gaussianity is detected in the cosmic
microwave background almost at 3$\sigma$ level \cite{Yadav:2007yy}
although this is not confirmed by the latest WMAP 5-year results
\cite{Komatsu:2008hk}.  (See also
Refs.~\cite{Hikage:2008gy,Slosar:2008hx,Curto:2008ym}.)  Further
observations of WMAP and future observations such as Planck can give
us more information on non-Gaussianity and serve to discriminate
different scenarios.

In most works done thus far on this topic, only one source of the
fluctuations has been considered. In general, however, since there
exist a lot of scalar fields in a supergravity or superstring theory,
different kinds of sources could contribute to the primordial
fluctuations simultaneously. Therefore, it is interesting to consider
models with mixed fluctuations from the inflaton and other sources. In
such mixed scenarios, not only non-Gaussianity but also other features
of the primordial fluctuations such as the amplitude, spectral index
and tensor modes are affected in comparison to the case where the
inflaton alone is the seed of fluctuations.  Current cosmological data
are already very precise regarding the scalar spectral index and the
tensor-to-scalar ratio and they can severely constrain models of
inflation. Some models of inflation are considered to have already
been excluded \cite{Komatsu:2008hk}. For example, chaotic inflation
with the higher order polynomial potential is excluded at more than
95\,\% confidence level. But, such constraints may be evaded by adding
another source of fluctuations. In fact, it is shown that some models
of inflation which are disfavored by the data can be liberated by
adding the curvaton contribution
\cite{Dimopoulos:2002kt,Endo:2003fr,Lazarides:2004we,Dimopoulos:2004yb,Rodriguez:2004yc,Langlois:2004nn,Moroi:2005kz,Moroi:2005np}.
In particular, in Ref.~\cite{Moroi:2005kz}, it has been studied in
what cases models of inflation can be relaxed by adding fluctuations
from the curvaton in some detail assuming some concrete inflation
models focusing on the spectral index and the tensor-to-scalar
ratio. Furthermore, in Ref.~\cite{Ichikawa:2008iq}, it was also
discussed in what cases/models large non-Gaussianity can be generated
satisfying the constraints on the scale dependence and tensor modes of
primordial fluctuations in the mixed scenario.
 
The modulated reheating scenario has been paid much attention recently
as interesting other source of fluctuations and large non-Gaussianity
but it is not so rigorously investigated as the curvaton mechanism in
particular as regards mixed scenarios with the inflaton.  Therefore,
in this paper, we consider the mixed models where fluctuations from
the inflaton and the modulated coupling can both contribute to the
present cosmic density fluctuations. Then, we study the effects of the
contribution of the modulated coupling on inflationary parameters such
as the spectral index and the tensor-to-scalar ratio and compare with
the WMAP 5-year results.  Furthermore, we also discuss in what cases
large non-Gaussianity can be generated satisfying the observational
constraints on the scale dependence and tensor modes of primordial
fluctuations in such a mixed scenario.

The structure of this paper is as follows.  In the next section, we
will give the expressions of the decay rate of the inflaton into
radiation for various types of interactions between the inflaton and
other fields and for various inflaton potentials, which will be used
in the subsequent sections.  In section~\ref{sec:Ne}, we will provide
the $e$-folding number from the time when the current cosmological
scales crossed the Hubble horizon during inflation to the time after
the inflaton decays, which enables us to evaluate the curvature
perturbations generated in this scenario.  Then we give the
expressions for the scalar spectral index, tensor-to-scalar ratio and
three non-linearity parameters in section~\ref{sec:observables}.  With
the formalism summarized in section~\ref{sec:observables}, we work on
some specific inflation models to compare the predictions of these
inflationary parameters with observations, paying particular attention
to in what cases non-Gaussianity can be very large.  For inflation
models which are considered to have already been excluded by the data,
we also give discussions in what case the contribution from the
modulated reheating can liberate the model.  The final section is
devoted to the conclusion and summary of this paper.

%%%%%%%%%%%%%%%%%%%%%%%%%%%%
\section{Decay rate of the inflaton}
%%%%%%%%%%%%%%%%%%%%%%%%%%%%

After inflation, the inflaton oscillates around the minimum of its
potential and the period of the oscillations is much shorter than the
expansion time characterized by the Hubble parameter.  The energy
density of the universe at that time is stored in the form of
oscillation energy of the inflaton.  Since the universe should become
radiation dominated before the time of big-bang nucleosynthesis, which
is required to be consistent with observations, the inflaton must
decay into radiation at some time after inflation.  Regarding
interactions between inflaton and radiation, we consider the following
Lagrangian,
\begin{eqnarray}
{\cal L}_{\rm int}\supset\ -\sum_a y_a(\sigma) \phi{\bar \psi}_a \psi_a -\sum_a M_a (\sigma) \phi \chi_a^2 -\sum_a h_a(\sigma) \phi^2 \chi_a^2,
\end{eqnarray}
where $\phi$ is the inflaton and $\chi_a$ and $\psi_a$ are scalar and
spinor fields which constitute radiation ($a$ represents the species
of the particles).  In the modulated reheating scenario, the coupling
constants $y_a,~M_a$ and $h_a$ are functions of a scalar field which
we denote as $\sigma$.  Although the number of such scalar fields is
not necessarily one, we consider only one modulus to avoid inessential
complexity.

The oscillations of the inflaton act as a periodically changing
external field on $\chi_a$ and $\psi_a$ fields.  This external field
creates $\chi_a$ or $\psi_a$ particles out of the vacuum.  Due to the
total energy density conservation, the energy of created particles are
compensated by the loss of the oscillation energy of the inflaton.
Hence this process can be regarded as the particle production from the
inflaton decay.

Let us suppose that the inflaton potential around the minimum can be
well approximated by a polynomial form as $V(\phi) \propto \phi^{2n}$
with $n$ being a positive integer.  Denoting the inflaton energy
density as $\rho_\phi$, the decay rate of the inflaton to the lowest
order in the coupling constants is given by
\begin{eqnarray}
\Gamma_\phi^{(n)}(\sigma)=\sum_a A_n \frac{y_a^2(\sigma)}{8\pi} m_\phi^{\rm eff}+\sum_a B_n \frac{M_a^2(\sigma)}{8\pi m_\phi^{\rm eff}}+\sum_a C_n \frac{h_a^2(\sigma)}{8\pi {(m_\phi^{\rm eff})}^3} \rho_\phi,  
\end{eqnarray}
where $m_\phi^{\rm eff}$ is the effective mass of the inflaton defined
by
\begin{eqnarray}
{(m_\phi^{\rm eff})}^2=V_{\phi \phi} \bigg |_{\phi={\bar \phi}},
\end{eqnarray}
with ${\bar \phi}$ being the amplitude of the oscillations. Here and
hereafter, the subscript $\phi$ of the potential $V$ represents the
derivative with respect to $\phi$.  $A_n,~B_n$ and $C_n$ are numerical
coefficients of ${\cal O}(1\sim 100)$.  The explicit values of these
are given in the appendix~\ref{sec:decay}.

%%%%%%%%%%%%%%%%%%%%%%%%%%%%%
\section{The number of $e$-folding}\label{sec:Ne}
%%%%%%%%%%%%%%%%%%%%%%%%%%%%%

In the following, the $e$-folding number especially plays an important
role in two aspects.  First of all, when we calculate observables such
as the amplitude, spectral index, non-Gaussianity and so on of the
curvature perturbations, we have to know when the cosmological scales
exited the Hubble horizon during inflation, i.e. the $e$-folding
number from the time of horizon crossing during inflation to the
present.  To this end, we need to know how the universe evolved from
the inflationary universe into the standard radiation dominated
universe, which obviously requires our knowledge of how the universe
is reheated by the inflaton decay.  Another aspect is that knowledge
of the $e$-folding number enables us to calculate the curvature
perturbations to any order in the perturbative expansion without
invoking complicated perturbed equations but using the so-called
$\delta N$ formalism
\cite{Starobinsky:1986fxa,Sasaki:1995aw,Sasaki:1998ug,Lyth:2004gb}.
All the information regarding various observables of the curvature
perturbations is contained in the $e$-folding number.  In this
section, we give an expression for the $e$-folding number from the
time $t_\ast$ when the cosmological scale crossed the horizon during
inflation to the time $t_f$ when the universe is completely reheated
by the inflaton decay.  After $t_f$, we assume that the universe
evolves according to the standard hot Big-Bang model.

Let us denote the $e$-folding number from $t_*$ to $t_f$ as $N
(t_f,t_*, \phi_*, \sigma_*)$.  Here $\phi_*$ and $\sigma_*$ are values
of $\phi$ and $\sigma$ at $t=t_*$.  For later convenience, we divide
$N(t_f, t_*, \phi_*, \sigma_*)$ into two parts as
\begin{equation}
\label{eq:N_tot}
N(t_f, t_*, \phi_*, \sigma_*) = N(t_{\rm end}, t_*, \phi_*)  + N(t_f, t_{\rm end}, \sigma_*),
\end{equation}
where $t_{\rm end}$ is the time at the end of inflation. Hence, on the
right-hand side, the first term represents the $e$-folding number
during inflation and the second term remaining one after the
inflation.  In writing this equation, we have implicitly assumed that
the mass and vacuum expectation value of the modulus are sufficiently
small and the background dynamics during inflation is completely
determined by the inflaton alone, which means that the first term on
the right-hand side is a function of $\phi_\ast$ only.  Meanwhile, the
second term depends only on $\sigma_\ast$.

By using the slow roll approximation, we write the first term as
\begin{equation}
\label{eq:N_inf}
N(t_{\rm end}, t_*, \phi_*)= \int_{t_\ast}^{t_{\rm end}} H dt  
\simeq  
- \frac{1}{M_{\rm pl}^2} \int_{\phi_\ast}^{\phi_{\rm end}}  \frac{V}{V_\phi} d \phi,
\end{equation}
where $\phi_\ast$ and $\phi_{\rm end}$ represent the scalar field
values at corresponding epochs.  To give a more concrete expression
for $N(t_{\rm end}, t_*, \phi_*)$, we need to specify the potential
for the inflaton.

Let us next consider the second term $N(t_f, t_{\rm end}, \sigma_*)$.
When the slow-roll conditions are violated, the inflaton starts to
oscillate around its minimum.  After several oscillations, the energy
density of the inflaton can be well approximated by that of a perfect
fluid with a constant equation of state.  If the potential around the
minimum is written as $V \propto \phi^{2n}$, its energy density
decreases as $\rho_\phi \propto a^{-6n/(n+1)}$.  For the discussion of
the curvature perturbations in the next section, we further decompose
the second term into two parts. Taking $t_c$ as a time after several
oscillations of the inflaton but well before the time of decay,
$N(t_f,t_{\rm end},\sigma_*)$ can be divided as
\begin{eqnarray}
N(t_f,t_{\rm end},\sigma_*)=N(t_c,t_{\rm end})+N(t_f,t_c,\sigma_*). \label{devide1}
\end{eqnarray}
Note that the first term on the right-hand side does not depend on
$\sigma_*$ as long as $t_c$ is taken to be sufficiently before the
time of the decay.  As will become clear later, $N(t_c,t_{\rm end})$
is irrelevant to the curvature perturbations.  However, we have to
take into account it when we calculate the epoch when the reference
scale at present, which is taken to be $k = 0.002~{\rm Mpc}^{-1}$ for
our analysis, exited the horizon during inflation.  In general, we
need a numerical calculation to evaluate $N(t_c,t_{\rm end})$ and do
not discuss this term further here.

\begin{figure}[t]
\begin{center}
\begin{tabular}{cc}
\resizebox{100mm}{!}{\includegraphics{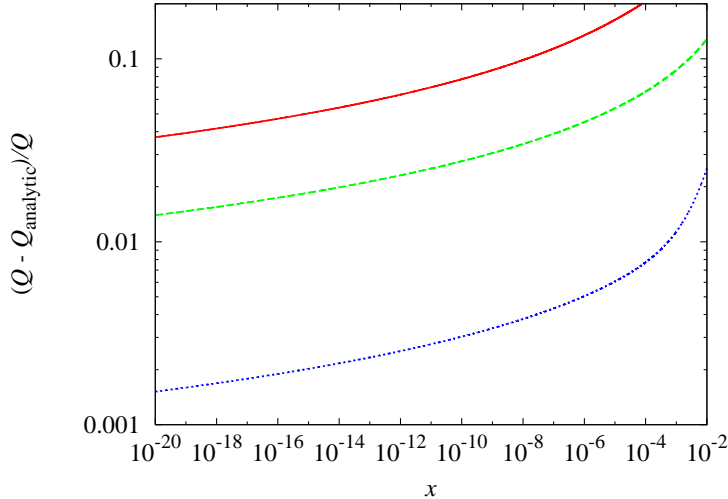}}
\end{tabular}
\caption{The relative error for the function $Q(x)$ between the one
  obtained by analytic and numerical methods for the cases with the
  interaction $-y\phi \bar{\psi}\psi$ (red solid line), $-M\phi
  \chi^2$ (green dashed line) and $-h\phi^2 \chi^2$ (blue dotted
  line).  Here we assumed $V \propto \phi^6$ for the inflaton
  potential.  For the cases with the quadratic potential $V \propto
  \phi^2$, the errors are smaller than those given in this figure. }
\label{fig:errorQ}
\end{center}
\end{figure}

Regarding $N(t_f,t_c,\sigma_*)$, for our discussion in the following,
we write it as
\begin{eqnarray}
&&N(t_f,t_c,\sigma_*)=\frac{1}{4} \log \frac{\rho_c}{\rho_f}+Q \left[ \Gamma_\phi(\sigma_*,t_c)/H_c \right], \label{e1} 
\end{eqnarray}
where $\rho_c$ and $\rho_f$ are total energy densities of the universe
at $t_c$ and $t_f$ respectively.  This equation should be understood
as a definition of the function $Q$.  Note that $Q$ depends on the
variable $x\equiv \Gamma(\sigma_*,t_c)/H_c$ alone, which can be
confirmed by a dimensional analysis of the background evolution
equations.  Since the dependence of $Q$ on $x$ differs depending on
the expansion law of the background space-time, i.e. the power $2n$
that determines the form of the inflaton potential, and also on the
dominant decay channel of the inflaton, we need to follow the
background evolution for each case.  In principle, we have to evaluate
it by numerical calculations.  However, we can make approximate
analytic estimates for $Q(x)$ when $x \ll 1$.  We found that $Q(x)$
can be well approximated with the form $Q(x)=a_0 \log x$ where $a_0$
is a numerical constant which depends on the interaction between the
inflaton and matter and the potential for the inflaton.  In
Table~\ref{table}, the values of $a_0$ are listed for three different
powers for the inflaton potential and three different interactions,
which are obtained by analytic estimates where the sudden-decay
approximation is adopted.  The derivation of these analytic values is
given in Appendix \ref{sec:app_Q}.  We have also calculated them
numerically and found that the differences between analytic estimate
and numerical one are at most $10$ \% for $x<10^{-6}$, which can be
seen from Fig.~\ref{fig:errorQ} where the relative error for the case
with the inflation potential being $V \propto \phi^{6}$ and three
types of interaction listed in Table~\ref{table}.

\begin{table}[t]
\begin{center}
\begin{tabular} {|c|c|c|c|}\hline
${\cal L}_{\rm int}$ &$n=1$&$n=2$&$n=3$ \\ \hline
$-y\phi {\bar \psi} \psi$ & $-\frac{1}{6}$ &0& $\frac{1}{6}$ \\ \hline
$-M\phi \chi \chi$ & $-\frac{1}{6}$ &0& $\frac{1}{30}$ \\ \hline
$-h\phi^2 \chi^2$ & --- &0& $\frac{1}{18}$ \\ \hline
\end{tabular}
\caption{We list the analytic values of $a_0$ defined by $Q(x)=a_0
  \log x$ for various inflaton potential $V(\phi) \propto \phi^{2n}$
  and dominant interactions.  Notice that these values are in good
  agreement with numerically obtained ones, in particular for small
  values of $x$.  If $n=1$, the inflaton cannot decay only with the
  four-point interaction $-h\phi^2 \chi^2$.}
\label{table}
\end{center}
\end{table}

In the following analysis, we consider the cases with $n=1,2$ and $3$.
Here we briefly discuss the tendency of $Q$ for these cases in order.
When $n=1$, the inflaton potential can be written as
$V(\phi)=\frac{m^2}{2}\phi^2$.  In this case, we have $m_\phi^{\rm
  eff}=m$ and the decay rate becomes independent of time for the
interactions such as $-y \phi {\bar \psi}\psi$ and $-M\phi \chi^2$.
Thus we obtain the same value of $Q$ for these interactions.
Meanwhile, if the dominant decay occurs through the four-point
interaction, the inflaton cannot decay completely into radiation
simply because the decay rate which is proportional to the Hubble
parameter squared decreases faster than the Hubble parameter.  Thus in
this case, the universe cannot be reheated.  Hence we do not consider
a four-point interaction case for the quadratic potential.  When
$n=2$, $\rho_\phi$ behaves the same as the energy density of
radiation.  Hence the universe expands in exactly the same way as the
background space-time even if the decay rate of the inflaton slightly
deviates from the background value\footnote{
  This is not true if preheating occurs.  In
  Ref.~\cite{Suyama:2006rk}, it was shown that the universe can evolve
  differently at different locations separated by the super-horizon
  distance during preheating because of the persisting isocurvature
  perturbations at the end of inflation.
}.  This means that $Q=0$ for any interactions.  For the case with
$n=3$, since the effective mass $m_\phi^{\rm eff}$ changes in time,
the decay rate evolves differently for different types of a dominant
interaction, which gives a different form of $Q$.

Also note that since the coefficient of $\log x$ in $Q(x)$ is negative
(positive) for $n=1~ (n=3)$, the function $Q(x)$ itself becomes
positive (negative) for $n=1~(n=3)$ when $x \le 1$.  From the
definition of $Q$ (see Eq.~(\ref{e1})), $Q$ represents the deviation
of the $e$-folding number from the radiation dominated universe.
Because energy density of dust decays more slowly than the radiation
energy density, the more dust gives the more $e$-folding number until
the total energy density decreases to a fixed value, which means the
positive $Q$.  Meanwhile, because the energy density of the inflaton
oscillating in the sextic potential decays as $a^{-9/2}$, it decreases
more rapidly than the energy density of radiation.  Hence $Q$ becomes
negative for the sextic potential.

In closing this section, we write down the expression for
$N(t_f,t_*,\phi_*,\sigma_*)$:
\begin{eqnarray}
N(t_f,t_*,\phi_*,\sigma_*)=-\frac{1}{M_{\rm pl}^2}\int_{\phi_*}^{\phi_{\rm end}} d\phi ~\frac{V}{V_\phi}
       +\frac{1}{4} \log \frac{\rho_c}{\rho_f}+N(t_c,t_{\rm end})
       +Q \left[ \Gamma_\phi(\sigma_*)/H_0 \right], \label{efold1}
\end{eqnarray}
which forms the basis in calculating various observables in the
subsequent sections.

%%%%%%%%%%%%%%%%%%%%%%%%%%%%
\section{Observables}\label{sec:observables}
%%%%%%%%%%%%%%%%%%%%%%%%%%%%

\subsection{Curvature perturbation}

To calculate the curvature perturbation, we make use of the $\delta N$
formalism
\cite{Starobinsky:1986fxa,Sasaki:1995aw,Sasaki:1998ug,Lyth:2004gb}.
In this formalism, the primordial curvature perturbation $\zeta$ on
the uniform energy density hypersurface at the time $t=t_f$ is given
by differentiating the $e$-folding number $N(t_f,t_*,\phi_*,\sigma_*)$
in Eq.~(\ref{efold1}) with respect to $\phi_*$ and $\sigma_*$,
\begin{eqnarray}
\zeta \approx \frac{1}{M_{\rm pl}^2}\frac{V}{V_\phi} \delta \phi_*+\frac{1}{2M_{\rm pl}^2} \left( 1-\frac{V V_{\phi \phi}}{V_\phi^2} \right) \delta \phi_*^2+\frac{1}{6M_{\rm pl}^2} \left( -\frac{V_{\phi \phi}}{V_\phi}-\frac{V V_{\phi \phi \phi}}{V_\phi^2}+2 \frac{V V_{\phi \phi}^2}{V_\phi^3} \right) \delta \phi_*^3 \nonumber \\
+Q_\sigma \delta \sigma_*+\frac{1}{2} Q_{\sigma \sigma} \delta \sigma_\ast^2+\frac{1}{6} Q_{\sigma \sigma \sigma} \delta \sigma_*^3, \label{eq:zeta_expansion}
\end{eqnarray}
where $\delta \phi_*$ and $\delta \sigma_*$ are the perturbations of
$\phi_*$ and $\sigma_*$ on the flat slicing.  For the purpose of this
paper, we include the terms up to cubic order in the perturbations of
scalar fields.  If we choose $t_f$ well after the reheating, then
$\zeta (t_f)$ gives primordial adiabatic perturbations.

Using $x=\Gamma_\phi (\sigma_\ast,t_c)/H_c$, the derivatives of $Q$
with respect to $\sigma$ can be expressed using $\Gamma_\phi$ as
\begin{eqnarray}
&&Q_\sigma=x Q'(x) \frac{\Gamma_\sigma}{\Gamma}
=A(x) \frac{\Gamma_\sigma}{\Gamma}, \\
&&Q_{\sigma \sigma}=x Q'(x) \frac{\Gamma_{\sigma\sigma}}{\Gamma} 
+x^2 Q''(x) \frac{\Gamma_\sigma^2}{\Gamma^2}
=A(x)\frac{\Gamma_{\sigma\sigma}}{\Gamma} 
+B(x)\frac{\Gamma_\sigma^2}{\Gamma^2} ,\\
&&Q_{\sigma \sigma \sigma}=x Q'(x) \frac{\Gamma_{\sigma\sigma\sigma}}{\Gamma} 
+3x^2 Q''(x) \frac{\Gamma_\sigma \Gamma_{\sigma\sigma}}{\Gamma^2}
+x^3 Q'''(x) \frac{\Gamma_\sigma^3}{\Gamma^3} \nonumber \\
&& \hspace{10mm}
= 
A(x)\frac{\Gamma_{\sigma\sigma\sigma}}{\Gamma} 
+ 3B(x)  \frac{\Gamma_\sigma \Gamma_{\sigma\sigma}}{\Gamma^2}
+C(x) \frac{\Gamma_\sigma^3}{\Gamma^3}. 
\label{second1.52}
\end{eqnarray}
Here we defined the function as $A(x) \equiv x Q(x)^\prime, B(x) =
x^2Q(x)^{\prime \prime}$ and $C(x) = x^3 Q(x)^{\prime\prime\prime}$
and denote the derivatives of $\Gamma_\phi$ with respect to
$\sigma_\ast$ as $\Gamma_\sigma$, $\Gamma_{\sigma\sigma}$, and
$\Gamma_{\sigma\sigma\sigma}$.  When $x \ll 1$, which we consider in
the followings, these functions become almost independent of $x$,
namely $Q$ can be well approximated by $Q(x) =a_0\log x$ (see Table
\ref{table}).  In this case, the functions defined above become
constant and can be written as $A = a_0, B = -a_0$ and $C = 2 a_0$.

\subsection{Power spectrum}

The power spectrum $P_\zeta$ of the curvature perturbations is defined
by
\begin{eqnarray}
\langle \zeta_{\vec k_1} \zeta_{\vec k_2} \rangle_c =
{(2\pi)}^3 P_\zeta (k_1) \delta ({\vec k_1}+{\vec k_2}),  \label{eq:power}
\end{eqnarray}
where $\langle \cdots \rangle_c$ means that we take connected parts of
$\langle \cdots \rangle$.

By using Eq.~\eqref{eq:zeta_expansion}, we can express ${\cal
  P}_\zeta(k) \equiv k^3 P_\zeta (k)/2\pi^2$ to the leading order in
$\delta \phi_*,~\delta \sigma_*$ as
\begin{equation}
{\cal P}_\zeta (k)= \frac{1}{2\epsilon} {\left( \frac{H_*}{2\pi M_{\rm pl}} \right)}^2 (1+R),  \label{eq:P_zeta} 
\end{equation}
where $\epsilon$ is a slow-roll parameter defined by,
\begin{eqnarray}
\epsilon \equiv \frac{M_{\rm pl}^2}{2} \frac{V_\phi^2}{V^2}.
\end{eqnarray}
Furthermore $R \equiv 2\epsilon A(x)^2 M_{\rm pl}^2
\Gamma_\sigma^2/\Gamma^2$ is the square of the ratio of the curvature
perturbation from modulated reheating to that from the inflaton, i.e.,
$\zeta_{\rm mod}^2 / \zeta_{\rm inf}^2$.  Thus the limit $R
\rightarrow 0$ ($R \rightarrow \infty$) corresponds to the case where
the curvature perturbation is sourced only by the inflaton (modulus)
fluctuations.  For discussion in the following, here we give the
definition of another slow-roll parameter $\eta$:
\begin{eqnarray}
\eta \equiv M_{\rm pl}^2\frac{V_{\phi \phi}}{V}.
\end{eqnarray}

\subsection{Bispectrum}

The bispectrum $B_\zeta$ is defined by
\begin{eqnarray}
\langle \zeta_{\vec k_1} \zeta_{\vec k_2} \zeta_{\vec k_3} \rangle_c=
{(2\pi)}^3 B_\zeta (k_1,k_2,k_3) \delta ({\vec k_1}+{\vec k_2}+{\vec k_3}). \label{eq:bi} 
\end{eqnarray}
During inflation, both $\phi$ and $\sigma$ are slowly-rolling.  It is
known that the non-Gaussianity of $\zeta$ coming from the intrinsic
non-Gaussianities of $\delta \phi_*$ and $\delta \sigma_*$ is far
below the observational sensitivity.  Hence we can treat $\delta
\phi_\ast$ and $\delta \sigma_\ast$ as uncorrelated Gaussian random
fields with the same amplitude.

If we parameterize $B_\zeta$ by the dimensionless parameter $f_{\rm
  NL}$ by
\begin{eqnarray}
B_\zeta (k_1,k_2,k_3)=\frac{6}{5} f_{\rm NL} 
\left( 
P_\zeta (k_1) P_\zeta (k_2) 
+ P_\zeta (k_2) P_\zeta (k_3) 
+ P_\zeta (k_3) P_\zeta (k_1)
\right), \label{eq:def_f_NL}
\end{eqnarray}
then $f_{\rm NL}$ can be written as
\begin{eqnarray}
\frac{6}{5}f_{\rm NL} = 
\frac{R^2}{A(x) {(1+R)}^2} 
\left( \frac{B(x)}{A(x)}+\frac{\Gamma \Gamma_{\sigma \sigma}}{\Gamma_\sigma^2} 
\right)+{\cal O}(\epsilon,\eta). \label{f1}
\end{eqnarray}
The first term $A(x)^{-2}B(x)R^2/{(1+R)}^2$ which is independent of
how the decay rate $\Gamma$ depends on $\sigma$ represents the
non-Gaussianity coming from the non-linearity between $\zeta$ and
$\delta \Gamma$.  Because the function $R^2/{(1+R)}^2$ is suppressed
by $R^2$ for $R\ll 1$ and approaches $1$ for $R \gg 1$, the magnitude
of the first term is at most $|A^{-2} B|={\cal O}(1\sim 10)$.  For
example, if the potential is given by a quadratic term, we have
$|A^{-2}B| \simeq 6$ (see Table~\ref{table}), which yields $f_{\rm
  NL}=5$.  The non-linearity between $\zeta$ and $\delta \Gamma$ gives
the positive (negative) $f_{\rm NL}$ for quadratic (sextic) inflaton
potential.

On the other hand, the second term which represents the
non-Gaussianity coming from the non-linearity between $\Gamma$ and
$\sigma$ depends on the detailed form of $\Gamma (\sigma)$.  We see
that very large non-Gaussianity $|f_{\rm NL}| \gg 1$ can be generated
only when $|\Gamma \Gamma_{\sigma \sigma}/\Gamma_\sigma^2| \gg 1$ is
satisfied.

\subsection{Trispectrum}

\begin{figure}[t]
\begin{center}
\begin{tabular}{cc}
\resizebox{100mm}{!}{\includegraphics{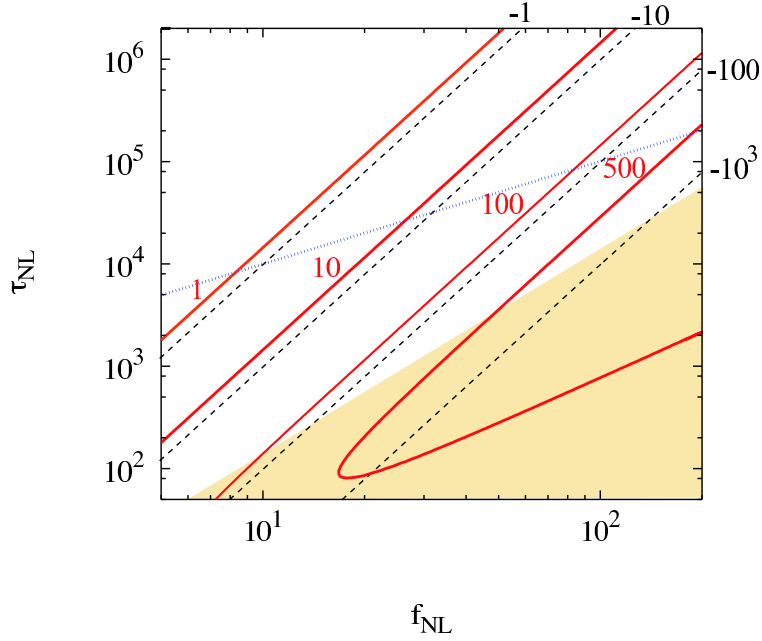}}
\end{tabular}
\caption{Consistency relations among three non-linearity parameters
  are shown for the case with modulated reheating corresponding to
  Eq.~\eqref{eq:consistency} (red solid line) and the curvaton model
  for the case where the curvaton decays before it dominates the
  universe corresponding to Eq.~(77) in Ref~\cite{Ichikawa:2008iq}
  (black dashed line).  In this figure, the consistency relation is
  presented as contours of $g_{\rm NL}$ in the $f_{\rm
    NL}$--$\tau_{\rm NL}$ plane.  The line for models with
  ``ungaussiton" \cite{Suyama:2008nt} is also shown (blue dotted
  line). For this model, the relation between $f_{\rm NL}$ and
  $\tau_{\rm NL}$ is given, thus it is irrelevant to the value of
  $g_{\rm NL}$. Notice that the inequality Eq.~\eqref{eq:inequality}
  should hold for the above mentioned scenarios. Thus we also show a
  region violating this inequality with shade.}
\label{fig:consistency}
\end{center}
\end{figure}

The trispectrum $T_\zeta$ is defined by
\begin{eqnarray}
\langle \zeta_{\vec k_1} \zeta_{\vec k_2} \zeta_{\vec k_3} \zeta_{\vec k_4} \rangle_c =
{(2\pi)}^3 T_\zeta (k_1,k_2,k_3,k_4) \delta ({\vec k_1}+{\vec k_2}+{\vec k_3}+{\vec k_4}). \label{eq:tri}
\end{eqnarray}

If we parameterize $T_\zeta$ by the two dimensionless parameters
$\tau_{\rm NL}$ and $g_{\rm NL}$ as
\begin{eqnarray}
T_\zeta (k_1,k_2,k_3,k_4)
&=&
\tau_{\rm NL} \left( 
P_\zeta(k_{13}) P_\zeta (k_3) P_\zeta (k_4)+11~{\rm perms.} 
\right) \nonumber \\
&&
+ \frac{54}{25} g_{\rm NL} \left( P_\zeta (k_2) P_\zeta (k_3) P_\zeta (k_4)
+3~{\rm perms.} \right),
\label{eq:def_tau_g_NL} 
\end{eqnarray}
then these parameters are given by
\begin{eqnarray}
&&\tau_{\rm NL}
=
\frac{R^3}{A(x)^2 {(1+R)}^3} {\left(  \left( \frac{B(x)}{A(x)} \right)^2
-\frac{\Gamma \Gamma_{\sigma \sigma}}{\Gamma_\sigma^2} \right)}^2+{\cal O}(\epsilon^2,\epsilon \eta, \eta^2) \nonumber \\
&&\hspace{7mm}=\frac{36(1+R)}{25R} f_{\rm NL}^2+{\cal O}(\epsilon^2,\epsilon \eta, \eta^2), \label{tau1} \\
&&\frac{54}{25}g_{\rm NL}
=
\frac{R^3}{A(x)^2 {(1+R)}^3} 
\left(
 \frac{C(x)}{A(x)} 
 + \frac{3 B(x)}{A(x)} \frac{ \Gamma \Gamma_{\sigma \sigma}}{\Gamma_\sigma^2}
 +\frac{\Gamma^2 \Gamma_{\sigma \sigma \sigma}}{\Gamma_\sigma^3} \right)+{\cal O}(\epsilon \eta, \xi^2, \eta^2). \label{g1}
\end{eqnarray}

From Eqs.~(\ref{f1}) and (\ref{tau1}), neglecting the terms of the
slow-roll order, we find that
\begin{equation}
\label{eq:inequality}
\tau_{\rm NL} \ge \frac{36}{25} f_{\rm NL}^2.
\end{equation}
In Ref.~\cite{Suyama:2007bg}, it is shown that this inequality holds
in more general setting, i.e. for any cases where the leading
non-Gaussianity comes from super-horizon evolution.

As for $g_{\rm NL}$, the non-linearity between $\zeta$ and $\delta
\Gamma$ always gives positive values for $g_{\rm NL}$ irrespective of
the inflaton potential and the dominant interaction.  When the
non-linearity between $\delta \Gamma$ and $\delta \sigma$ cannot be
neglected, $g_{\rm NL}$ can be negative.

If the $\Gamma_{\sigma \sigma \sigma}$ term in $g_{\rm NL}$ can be
neglected and $x\ll 1$, all the three non-linear parameters are
functions of $R$ and $\Gamma \Gamma_{\sigma \sigma}/\Gamma_\sigma^2$.
This means that we have a universal relation among the non-linearity
parameters independent of $R$ and $\Gamma \Gamma_{\sigma
  \sigma}/\Gamma_\sigma^2$.  The explicit form of such a relation can
be written as
\begin{eqnarray}
\label{eq:consistency}
g_{\rm NL}=-\frac{864}{625 a_0^2} \frac{f_{\rm NL}^6}{\tau_{\rm NL}^3}
-\frac{12}{5a_0} \frac{f_{\rm NL}^3}{\tau_{\rm NL}}.
\end{eqnarray}
In particular, if the potential around the minimum is quadratic, then
we have $a_0=-\frac{1}{6}$ and the relation above reduces to
\begin{eqnarray}
\label{eq:consistency2}
g_{\rm NL}=-\frac{31104}{625} \frac{f_{\rm NL}^6}{\tau_{\rm NL}^3}+\frac{72}{5} \frac{f_{\rm NL}^3}{\tau_{\rm NL}}. \label{cons}
\end{eqnarray}
By checking this consistency relation from observations, we can
discriminate this scenario from other ones that also generate large
non-Gaussianity.  In Fig.~\ref{fig:consistency}, we show the contours
of $g_{\rm NL}$ as a function of $f_{\rm NL}$ and $\tau_{\rm NL}$
given in Eq.~(\ref{cons}).  For comparison, we also show the relation
among the non-linearity parameters which holds for other scenarios
which can generate large non-Gaussianity such as models with mixed
inflaton and curvaton for the case with a large positive $f_{\rm NL}$
\cite{Ichikawa:2008iq} and ``ungaussiton" \cite{Suyama:2008nt}.
Interestingly, while the curvaton gives negative $g_{\rm NL}$, the
modulated reheating gives positive one (if the potential is quadratic)
if $f_{\rm NL}$ is large positive.  Hence just the determination of
the sign of $g_{\rm NL}$ enables us to discriminate these two
competitive models in this case.  We also plot the relation between
$f_{\rm NL}$ and $\tau_{\rm NL}$ in the ``ungaussiton" model where the
relation between $\tau_{\rm NL}$ and $f_{\rm NL}$ is given
irrespective of $g_{\rm NL}$.  Thus only one line is drawn for this
model.

%%%%%%%%%%%%
\subsection{The scalar spectral index and the tensor-to-scalar ratio}
%%%%%%%%%%%%

In this subsection, we will give the expressions for the scalar
spectral index and tensor-to-scalar ratio. First of all, the scalar
spectral index $n_s$ is given by
\begin{eqnarray}
n_s-1 &\equiv& \frac{d \log {\cal P}_\zeta}{d \log k} \bigg |_{k=aH} \nonumber \\
&= & -2\epsilon-\frac{4\epsilon-2\eta}{1+R}. \label{ns1}
\end{eqnarray}

During inflation, the tensor mode is also generated and its power
spectrum can be written as
\begin{eqnarray}
{\cal P}_T = 8 {\left( \frac{H}{2\pi M_{\rm pl}} \right)}^2.
\end{eqnarray}
To characterize the size of the tensor mode, the tensor-to-scalar
ratio is usually used, which is given by
\begin{eqnarray}
r \equiv \frac{ {\cal P}_T }{ {\cal P}_\zeta}=\frac{16 \epsilon}{1+R}.
\end{eqnarray}

%%%%%%%%%%%%%%%%%%%%%%%%%%%%%%
\section{Inflationary predictions and comparison with observations}\label{sec:obs}
%%%%%%%%%%%%%%%%%%%%%%%%%%%%%%

In this section, we discuss the five inflationary parameters, the
scalar spectral index, tensor-to-scalar ratio and three non-linearity
parameters $(n_s,r,f_{\rm NL},\tau_{\rm NL},g_{\rm NL})$ in models
where fluctuations from modulated coupling and inflaton can both
contribute to the primordial curvature perturbations.  With regard to
$n_s$ and $r$, current cosmological observations give severe
constraints on these quantities, thus we also compare the predictions
for $n_s$ and $r$ with the WMAP 5-year data
\cite{Komatsu:2008hk,Dunkley:2008ie}.  For the purpose of this paper,
we consider the chaotic inflation models with some polynomials for
definiteness.  However, before we discuss its inflationary
predictions, here we make some comments on the cases with other
inflation models in a mixed scenario.  If we take the new inflation
models, its effects are expected to be very small since the new
inflation models usually give a negligibly small value of the
slow-roll parameter $\epsilon$. Since $\epsilon$ appears in $R =
2\epsilon A(x)^2 M_{\rm pl}^2 \Gamma_\sigma^2/\Gamma^2$ which
represents the size of the contribution from fluctuations from
modulated reheating, $R$ becomes very small when $\epsilon$ is
negligibly small.  This fact has already been pointed out in
Ref.~\cite{Ichikawa:2008iq} for a mixed model of the inflaton and
curvaton.  Hence, it is unlikely that fluctuations from modulated
reheating dominate the total curvature perturbations in the new
inflation model, in which we recover the usual standard formula for
the inflationary parameters and the effects of modulated reheating can
be negligible.  As another possible inflation model, we can also
assume the hybrid inflation model. However, as is also discussed in
Ref.~\cite{Ichikawa:2008iq}, the case of hybrid inflation leads to the
similar result as that of chaotic inflation. Thus we omit them in this
paper. Regarding the interactions between the inflaton and matter, we
adopt the Yukawa interactions, just to be concrete.  We can
straightforwardly analyze the case with other interactions by simply
replacing the numerical factor $A(x), B(x)$ and $C(x)$ appearing in
the expression with the corresponding values (see Table \ref{table}).

Even after we fix the form of the inflaton potential and the
interactions for reheating, we have still a degree of freedom of how
the coupling constant $y$ depends on the modulus $\sigma$.  In the
following, we assume that $y_a (\sigma)$ can be written as\footnote{
  Here we assume that there are no renormalizable interactions between
  $\sigma$ and $\psi_a,~\chi_a$ because such interactions will give the
  modulus a thermal mass much larger than the Hubble parameter, which
  suppresses fluctuations of modulus and spoils the modulated reheating
  scenario.
}. 
\begin{eqnarray}
y_a(\sigma)=y_a^0 \left( 1+\alpha_a \frac{\sigma}{M}+ \beta_a \frac{\sigma^2}{M^2}+\cdots \right),
\end{eqnarray}
where $\alpha_a$ and $\beta_a$ are ${\cal O}(1)$ coefficients. 
$M$ is some energy scale and 
we assume that $|\sigma| \ll M$.  
To be definite, we truncate the expansion of $y_a(\sigma)$ at the
second order in $\sigma$.  Then the decay rate can be also truncated
at the quadratic order in $\sigma$,
\begin{eqnarray}
\Gamma=\Gamma_0 \left( 1+\alpha \frac{\sigma}{M}+\beta \frac{\sigma^2}{M^2} \right), \label{eq:decay1}
\end{eqnarray}
where $\alpha$ and $\beta$ are also ${\cal O}(1)$ coefficients.

From Eq.~(\ref{eq:decay1}), we have
\begin{eqnarray}
\frac{\Gamma \Gamma_{\sigma \sigma}}{\Gamma_\sigma^2} \simeq \frac{2\beta}{ {\left( \alpha+\displaystyle\frac{2\beta \sigma}{M} \right)}^2}. \label{decay2}
\end{eqnarray}
If $\alpha$ is ${\cal O}(1)$, then this equation further reduces to
$\simeq 2\beta/\alpha^2$.  From Eq.~(\ref{f1}), the non-linearity
between $\Gamma$ and $\sigma$ gives $f_{\rm NL}\simeq
2\beta/(A(x)\alpha^2)$ for $R \gg 1$.  Since $|A(x)^{-1}|$ is ${\cal
  O}(1\sim 10)$, $|f_{\rm NL}|={\cal O}(10\sim 100)$ can be achieved
by setting $2\beta/\alpha^2$ to be ${\cal O}(10)$, which is quite
possible while satisfying $\alpha,\beta={\cal O}(1)$.  Notice that, to
obtain a large positive $f_{\rm NL}$, $\beta$ should be negative for
the quadratic potential since $A(x)<0$ for the potential.  For
example, if we take $(\alpha,\beta)=(0.5,-1)$ and assume the quadratic
potential and the Yukawa interaction for the inflaton in which $A(x)
\sim -1/6$, then we have $f_{\rm NL}=45$ for $R\gg 1$.  Since the
constraint on $f_{\rm NL}$ from WMAP 5-year data is given as $ - 9 <
f_{\rm NL} < 111$ \cite{Komatsu:2008hk}, this sort of possibilities
may be interesting.  Furthermore this demonstrates that the
non-linearity between $\Gamma$ and $\sigma$ can provide
non-Gaussianity of $\zeta$ much larger than that from the
non-linearity between $\zeta$ and $\delta \Gamma$.

Meanwhile, some symmetries may forbid the appearance of the linear
terms in $\sigma$ in $y_a (\sigma)$.  In this case, Eq.~(\ref{decay2})
becomes $\simeq M^2/(2\beta \sigma^2)$, where $f_{\rm NL}$ can be very
large because of $M/\sigma \gg 1$ when $\beta < 0$.  For an
illustrational purpose, we will consider two cases:
$(\alpha,\beta)=(0.3,-1.0)$ (case A) and $(\alpha,\beta)=(0.0,-1.0)$
(case B) in the following.

To compare the prediction for the primordial curvature fluctuations
and non-Gaussianity with observations, we need to specify when the
present cosmological scale exited the horizon during inflation.  Since
$k_\ast = a(t_\ast) H_\ast$ holds when the scale with the wave number
$k_\ast$ crossed the horizon, the reference scale $k_{\rm ref}$ where
we probe the primordial fluctuations at the present time is related to
that at the horizon crossing during inflation as
\begin{equation}
%\label{ }
\frac{k_{\rm ref}}{a_0 H_0} = \frac{a(t_*) H_\ast}{a_0 H_0},
\end{equation}
where $a_0$ and $H_0$ are the scale factor and the Hubble parameter at
present.  The ratio of $a_\ast$ to $a_0$ on the right-hand side can be
divided into several parts as,
\begin{equation}
%\label{ }
\frac{k_{\rm ref}}{a_0 H_0} = 
\frac{a(t_*)}{a(t_{\rm end})} 
\frac{a(t_{\rm end})}{a(t_f)}
\frac{a(t_f)}{a_0}
\frac{H_\ast}{H_0}.
\end{equation}
The definition of $t_{\rm end}$ and $t_f$ is given in
section~\ref{sec:Ne}.  By taking logarithm of both sides and using
Eqs.~(\ref{devide1}) and (\ref{e1}), the number of $e$-folding between
the time $t_\ast$ and $t_{\rm end}$, i.e., $N(t_{\rm end},t_*,\phi_*)$
can be written as
\begin{equation}
\label{eq:N_e}
N(t_{\rm end},t_*,\phi_*)
=-\log \frac{k_{\rm ref}}{a_0 H_0}-N(t_c,t_{\rm end})-\frac{1}{4} \log \frac{\rho_c}{\rho_f}-Q(x)
+ \log \frac{a_f}{a_0} + \log \frac{H_\ast}{H_0},
\end{equation}
with which we can determine the field value $\phi_\ast$ at the horizon
crossing.

For the reference scale, we take $k_{\rm ref} = 0.002~{\rm Mpc}^{-1}$
in the analysis.  As for the fifth term, we assume that no more
entropy is produced after the inflaton decays.  Thus this term can be
rewritten by using the conservation of the entropy density per
comoving volume.  Since the entropy density is given by $s = (2 \pi /
45) g_{\ast s}T^3$ with $g_{\ast s}$ being the total number of
effective massless degrees of freedom, we have the following relation,
\begin{equation}
%\label{ }
 \log \frac{a_f}{a_0} 
= \frac{1}{3} \log \frac{s_0}{s_f}=\frac{1}{3} \log \frac{g_{\ast s 0} T_0^3}{g_{\ast s f} T_f^3}.
\end{equation}
For $g_{\ast s f}$ at the time of $a_f$, we take $ g_{\ast s f}=100$.
With regard to other quantities, we assume $H(t_c)=10^{-2}H_{\rm end}$
and $x=10^{-8}$ in the following analyses.

\subsection{Chaotic inflation : $V(\phi)=\frac{m^2}{2} \phi^2$}

\begin{figure}[t]
\begin{center}
\resizebox{150mm}{!}{\includegraphics{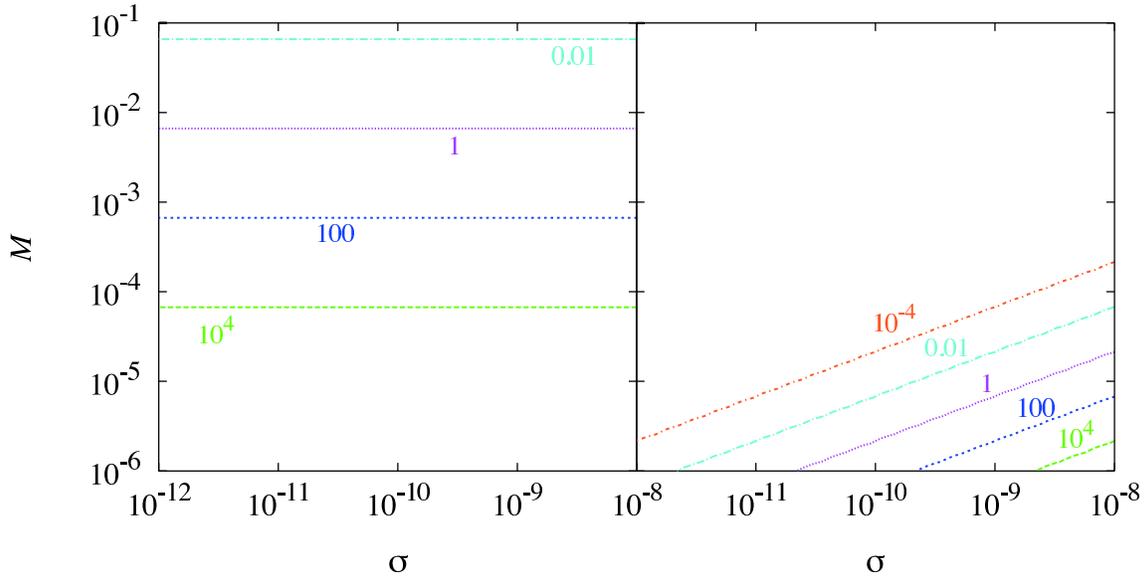}}
\caption{Contours of $R$ in the $\sigma$--$M$ plane for the chaotic
  inflation model with quadratic potential. Left and right panels are
  for case A ($\alpha = 0.3, \beta =-1$) and case B ($\alpha = 0,
  \beta =-1$), respectively.  In the figure, $M$ and $\sigma$ are
  shown in units of $M_{\rm pl}$.}
\label{chao2ratio}
\end{center}
\end{figure}

Now let us first consider chaotic inflation \cite{Linde:1983gd} with
the quadratic potential\footnote{
  This type of a simple polynomial potential can be realized in
  supergravity
  \cite{Goncharov:1983mw,Goncharov:1985ka,Murayama:1993xu,Kawasaki:2000yn,Kawasaki:2000ws,Kawano:2007gg,Kawano:2008iy}.
}.  Since we have fixed the values of the quantities which determine
the background evolution after the end of inflation as mentioned
above, remaining variables that we need to specify are some parameters
in the inflaton potential and the values of $\sigma$ and $M$ relevant
to fluctuations from modulated reheating, which appear in the decay
rate of the inflaton. Since the inflation is assumed to be driven
solely by the inflaton, the Hubble parameter during inflation is
controlled by the value of the parameters in the potential.  Thus the
primordial curvature fluctuations also depend on these parameters,
which means that the parameters in the inflaton potential can be fixed
by the WMAP normalization, i.e. by requiring that the total curvature
fluctuations are ${\cal O}(10^{-5})$.  In fact, the normalization
slightly depends on the spectral index and tensor-to-scalar ratio.
Hence we used the one given in Ref.~\cite{wmap5norm}\footnote{
The amplitude at $k=0.002~{\rm Mpc}^{-1}$ 
is given with $\delta_H^2 = (4/25) {\cal P_\zeta}$ as
\begin{eqnarray}
10^5 \delta_H^{\rm WMAP5} = 1.910 \times \frac{\exp \left[ (-0.724+0.533\, r)(1-n_s) \right]}{\sqrt{1+0.278\, r}}.
\end{eqnarray}
For the details, see Ref.~\cite{wmap5norm}.
}derived using the WMAP 5-year data in the following analyses to fix a
parameter in the potential for the inflaton.  For the quadratic case
of the chaotic inflation, we fix the value of $m$ by the WMAP
normalization. With regard to the spectral index $n_s$ and
tensor-to-scalar ratio $r$, the observation of WMAP 5-year also give
severe constraints on these quantities (see Fig.~3 in
Ref.~\cite{Komatsu:2008hk}).  When fluctuations from the inflaton
alone are responsible for the curvature fluctuations, in our setting
(fixing the parameters) where the number of $e$-folding during
inflation is $N_{\rm inf} \simeq 60$, these quantities are given as
$n_s = 0.967$ and $r=0.133$ which are allowed by WMAP 5-year data.
However, if fluctuations from the modulus fluctuations are included,
these predictions can be modified, which we will discuss in the
following.  To see in what cases such modifications are significant,
we show the contours of $R$ in the $\sigma$--$M$ plane in
Fig.~\ref{chao2ratio} for the case A (left panel) and B (right
panel). By looking at the figure, we can expect the parameter region
where the predictions for $n_s$ and $r$ are modified much.

\begin{figure}[t]
\begin{center}
\resizebox{150mm}{!}{\includegraphics{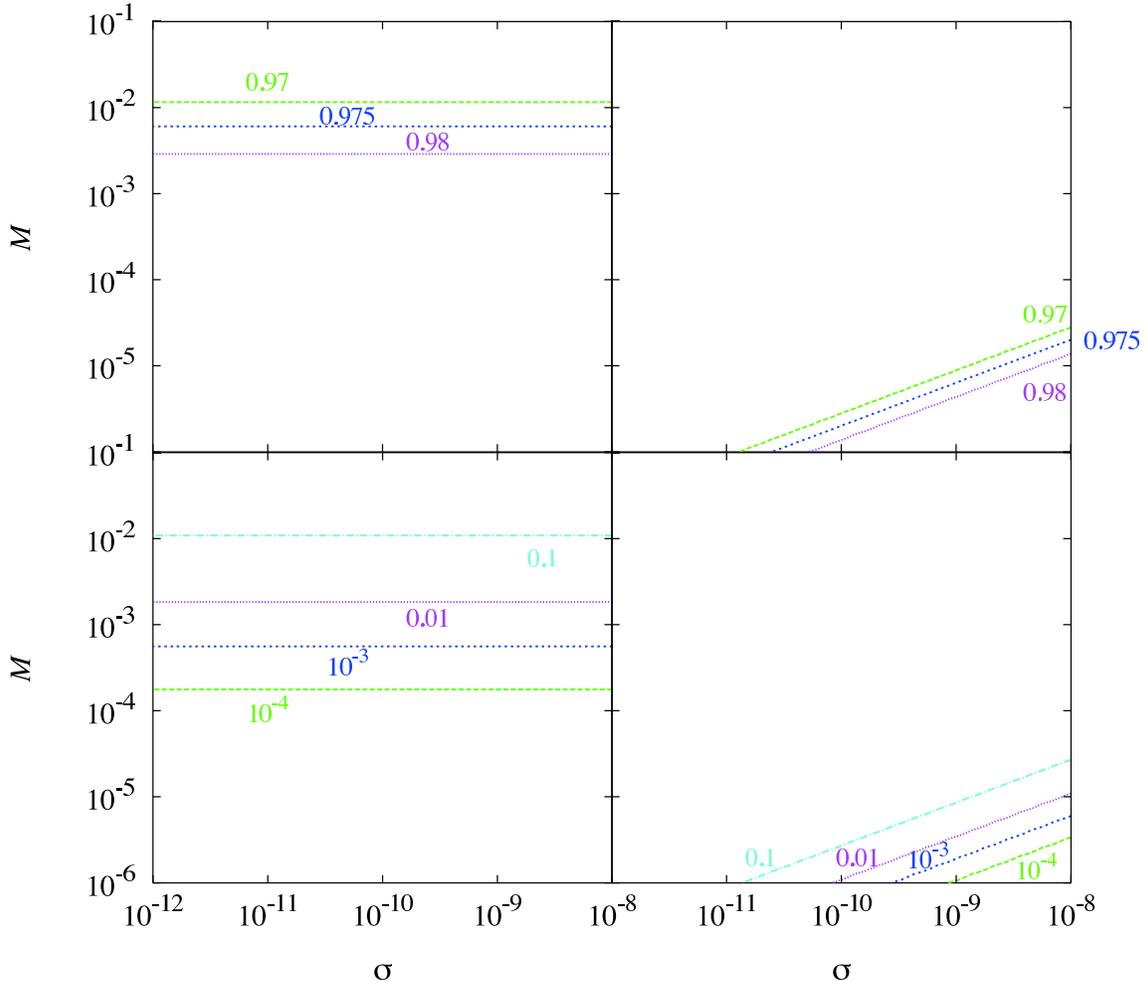}}
\caption{Contours of $n_s$ (top panels) and $r$ (bottom panels) in the
  $\sigma$--$M$ plane for the chaotic inflation model with quadratic
  potential. Left (right) panels are for case A (B).}
\label{chao2nsr}
\end{center}
\end{figure}

In the top panels of Fig.~\ref{chao2nsr}, we show contours of the
scalar spectral index $n_s$ in the $\sigma$--$M$ plane for the case A
(left panel) and B (right panel).  Let us first look at the left panel
(case A).  In this case, $\Gamma_\sigma/\Gamma$ is almost independent
of $\sigma$.  Hence $R$ is also independent of $\sigma$ and depends
only on $M$.  As can be seen from Eq.~(\ref{ns1}), the effect of
fluctuations from modulated reheating on $n_s$ appears only through
the parameter $R$.  This is the reason why the contours are parallel
to the $\sigma$-axis.  Furthermore, since $R \propto M^{-2}$, smaller
value of $M$ indicates larger $R$, i.e. larger contribution from the
modulus fluctuations to the total curvature perturbations.  When the
modulus contributions become dominant, the value of $n_s$ shifts from
$0.967$ to $0.983$.  For the case B (see the top right panel of
Fig.~\ref{chao2nsr}), $\Gamma_\sigma/\Gamma$ is proportional to
$\sigma/M^2$.  Hence the contours are parallel to the line
$M/\sigma^{1/2}={\rm const.}$, which can be seen from the figure.

Now we investigate how the tensor-to-scalar ratio is modified in the
mixed model. In the bottom panels of Fig.~\ref{chao2nsr}, we show
contours of the tensor-to-scalar ratio $r$ in the $\sigma$--$M$ plane.
Notice that as in the case of $n_s$, the effect of the modulus on $r$
appears only through $R$.  Hence the slope of the contours becomes the
same as those for $n_s$.  We see that when the modulus contributions
are dominant, $r$ becomes negligibly small.

\begin{figure}[!h]
\vspace{-1cm}
\begin{center}
\begin{tabular}{cc}
\resizebox{150mm}{!}{\includegraphics{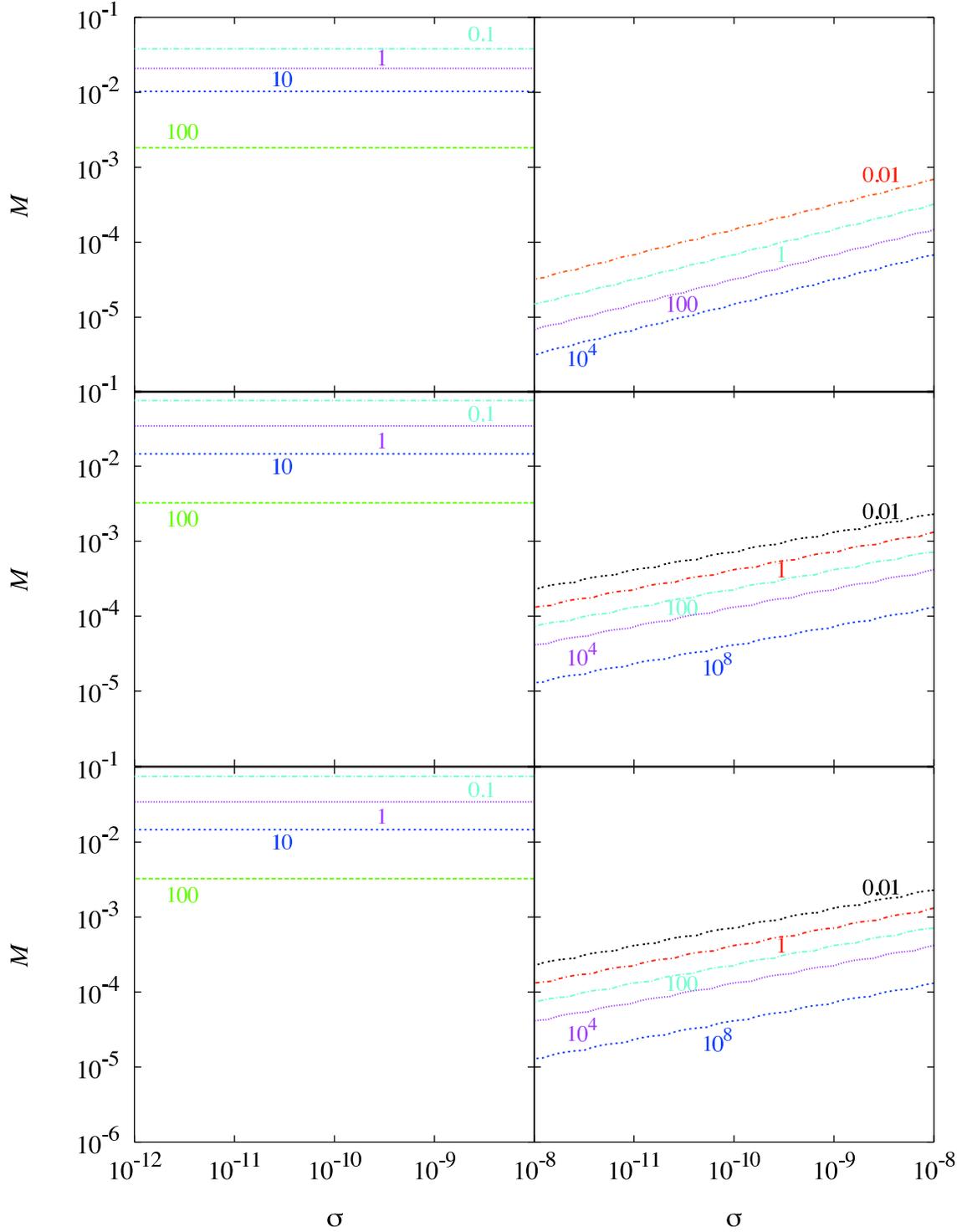}}
\end{tabular}
\caption{Contours of $f_{\rm NL}$ in the $\sigma$--$M$ plane for the
  chaotic inflation model with quadratic potential.  Left (right)
  panels are for case A (B).}
\label{chao2nl}
\end{center}
\end{figure}

Next we move on to the issues of non-Gaussianity.  For this purpose,
we show contours of non-linearity parameters $f_{\rm NL}, \tau_{\rm
  NL}$ and $g_{\rm NL}$ in the top, middle and bottom panels of
Fig.~\ref{chao2nl}, respectively.  Remember that, when the
fluctuations from the inflaton alone are assumed, these non-linearity
parameters are of the orders of the slow-roll parameters, which are
usually much less than unity. However, since fluctuations from
modulated reheating can give large non-Gaussianity, the non-linearity
parameters can be large even when the contribution from the modulus
fluctuations is subdominant in the curvature fluctuations.  For the
case A, we have $\Gamma \Gamma_{\sigma \sigma} /\Gamma_\sigma^2 \simeq
{\rm const.}$ (see Eq.~(\ref{decay2})).  Hence as in the cases for
$n_s$ and $r$, the value of $R$ alone determines the effects of the
modulus contributions on the non-linearity parameters.  Thus the
smaller $M$ gives the larger non-Gaussianity which can be seen from
the figures.  On the other hand, for the case B, the above mentioned
combination is $\Gamma\Gamma_{\sigma \sigma}/\Gamma_\sigma^2 \simeq
M^2/\sigma^2$.  Hence $\sigma$ and $R$ can both affect the
non-linearity parameters in this case.  If the former effect is
dominant, the slope of the contour becomes $1/2$.  On the other hand,
if the latter one is dominant, its slope becomes $1$ and we see that
for larger $R$ and smaller $M^2/\sigma^2$, we have larger
non-Gaussianity.  In fact, in Fig.~\ref{chao2nl}, contours
corresponding to the latter case are not shown for this model, since
the size of the non-linearity parameters become too large in such a
region for this model, thus we do not depict them here. However this
kind of behavior can be seen in Figs.~\ref{chao24nl} and \ref{chao6nl}
which are for other inflation models to be discussed in the following.

\begin{figure}[!h]
\begin{center}
\begin{tabular}{cc}
\resizebox{100mm}{!}{\includegraphics{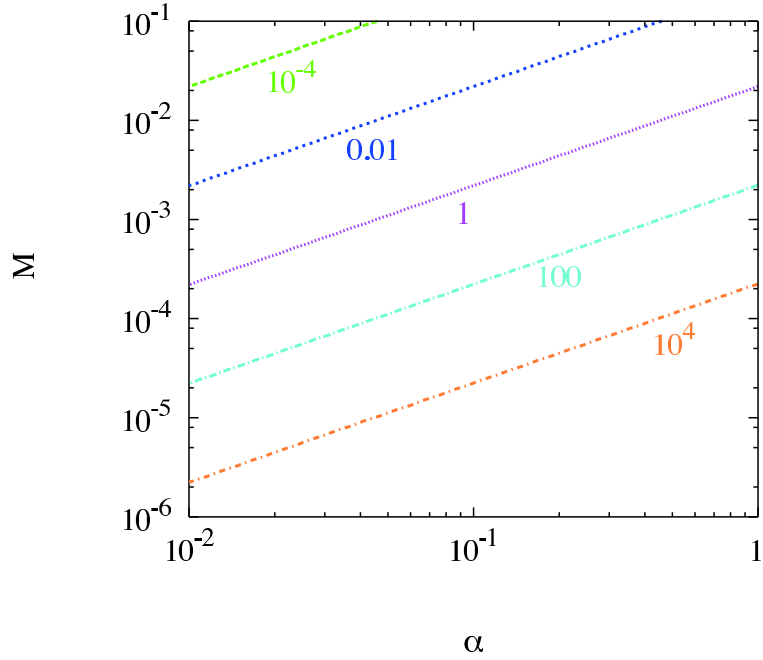}}
\end{tabular}
\caption{Contours of $R$ in the $\alpha$--$M$ plane for the chaotic
  inflation model with a quadratic potential. }
\label{chao2_alphaM_ratio}
\end{center}
\end{figure}

\begin{figure}[!h]
\begin{center}
\begin{tabular}{cc}
\resizebox{150mm}{!}{\includegraphics{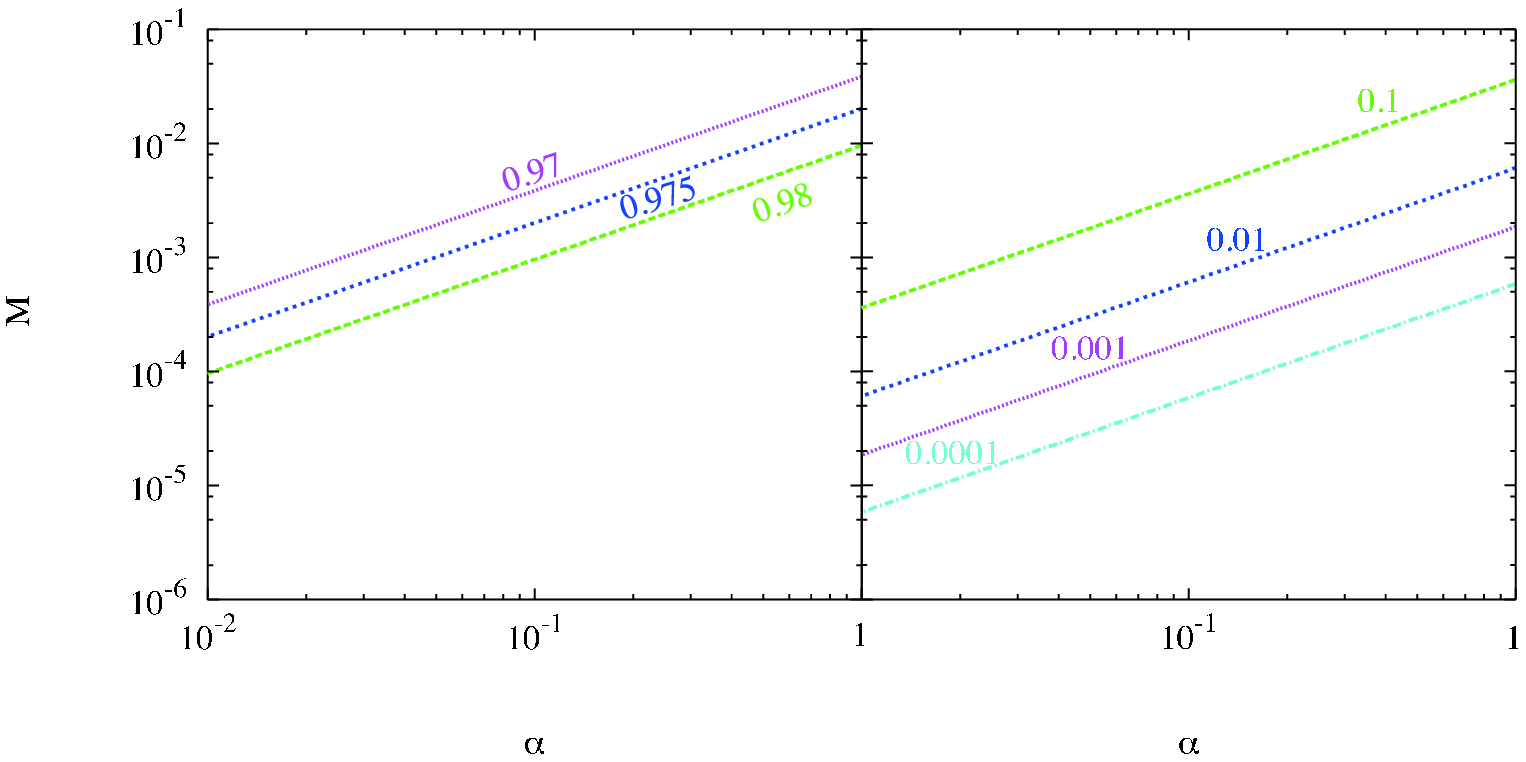}}
\end{tabular}
\caption{Contours of $n_s$ (left panel) and $r$ (right panel) in the
  $\alpha$--$M$ plane for the chaotic inflation model with a quadratic
  potential. }
\label{chao2_alphaM_nsr}
\end{center}
\end{figure}

For an illustrational purpose, we fix the values of $\alpha$ and
$\beta$. However, the predictions for the inflationary parameters also
depends on these values. Thus here we stop to discuss its
dependence. In Fig.~\ref{chao2_alphaM_ratio}, contours of the ratio
$R$ are shown in the $\alpha$--$M$ plane. Since the relative size
between $\alpha$ and $\beta$ is important to see the effects of the
modulated reheating, here we fix the value of $\beta$ and vary
$\alpha$.  Since the ratio $R$ is controlled by the combination
$\Gamma_\sigma^2 / \Gamma^2$, as $\alpha$ increases, $R$ also becomes
large for $M$ being fixed.  As discussed in the previous section,
$n_s$ and $r$ are also determined by the above combination, thus the
tendencies are the same as that of the ratio $R$, which can be seen in
Fig.~\ref{chao2_alphaM_nsr} where contours of $n_s$ and $r$ are
depicted.

\begin{figure}[!h]
\begin{center}
\begin{tabular}{cc}
\resizebox{150mm}{!}{\includegraphics{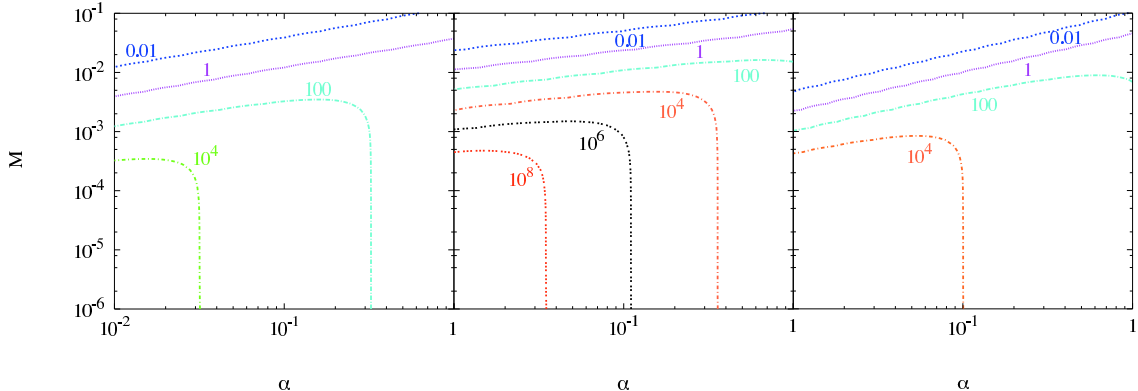}}
\end{tabular}
\caption{Contours of $f_{\rm NL}$ (left), $\tau_{\rm NL}$ (center) and
  $g_{\rm NL}$ (right) in the $\alpha$--$M$ plane for the chaotic
  inflation model with a quadratic potential.}
\label{chao2_alphaM_nl}
\end{center}
\end{figure}

However, if we look at plots of non-linearity parameters, which are
shown in Fig.~\ref{chao2_alphaM_nl}, the trends are different.  As
mentioned before, the non-linearity parameters are governed by the
combination $\Gamma \Gamma_{\sigma\sigma} / \Gamma_\sigma^2 \simeq 2
\beta/ (\alpha + 2\beta\sigma/M)^2$, when $\alpha$ is small, the
dependence of non-linearity parameters on $\alpha$ is also small. On
the other hand, when $\alpha$ is large, the size of the above
combination is determined by $\alpha$ with $\beta$ being fixed. Thus
the predictions are irrelevant to $M$ in this case, which can be read
off from the figure.  In the following, we discuss other types of the
inflaton potential and again present our results in the $\sigma$--$M$
plane fixing the values of $\alpha$ .  However, the trend discussed
here also applies to those cases.

%%%%%%%%%%%%%%%%%%%%%%%%%%%%%%%%%%%%%
\subsection{Chaotic inflation : 
$V(\phi) = \frac{\lambda}{4}\phi^4 + \frac{m^2}{2} \phi^2 $
}

\begin{figure}[!h]
%\vspace{-1.2cm}
\begin{center}
\begin{tabular}{cc}
\resizebox{150mm}{!}{\includegraphics{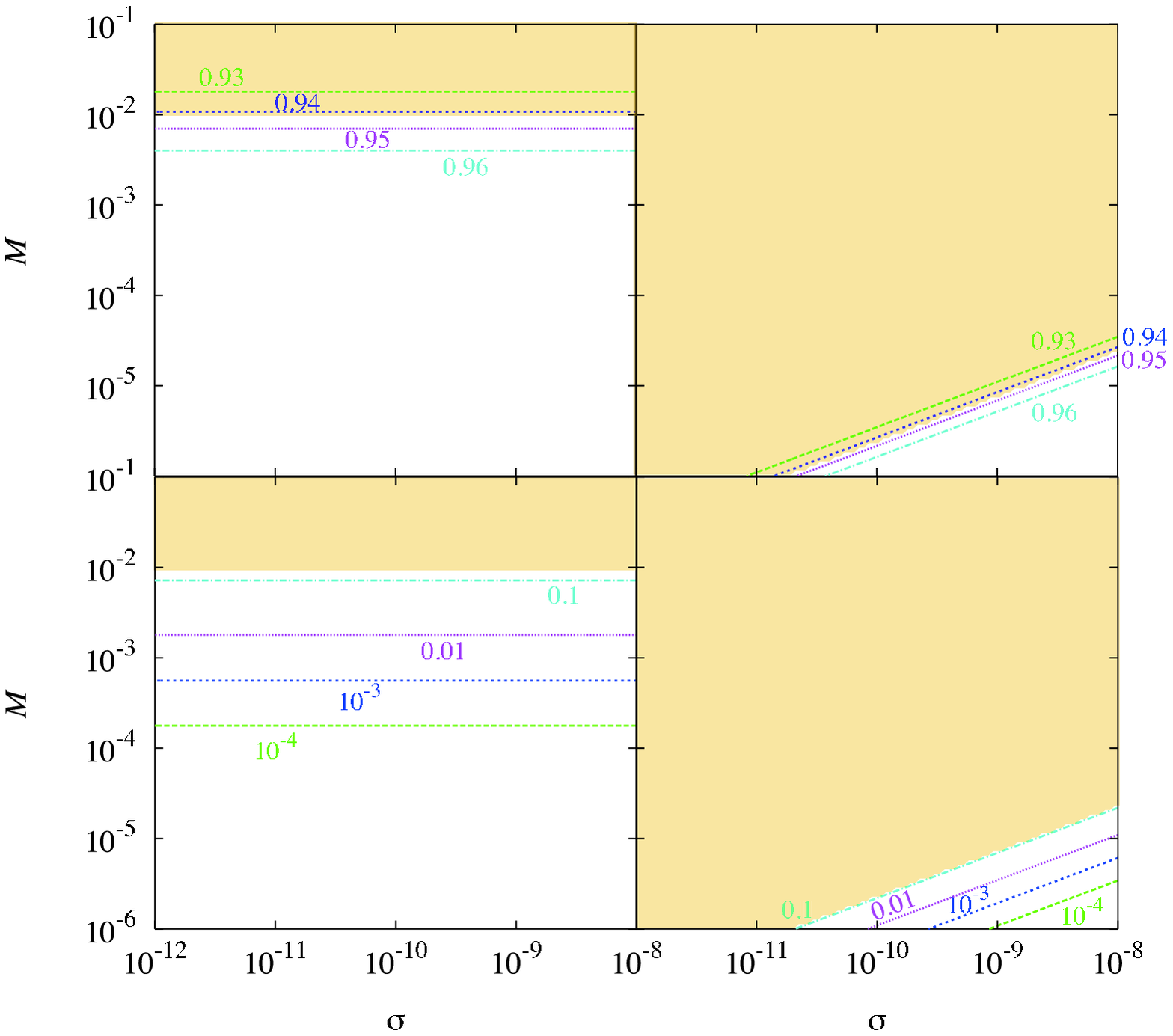}}
\end{tabular}
\caption{Contours of $n_s$ (top panels) and $r$ (bottom panels) in the
  $\sigma$--$M$ plane for the chaotic inflation model with quadratic
  and quartic potential.  Left (right) panels are for case A
  (B). Shaded region is excluded by WMAP5.}
\label{chao24nsr}
\end{center}
\end{figure}

\begin{figure}[!h]
\vspace{-1.2cm}
\begin{center}
\begin{tabular}{cc}
\resizebox{150mm}{!}{\includegraphics{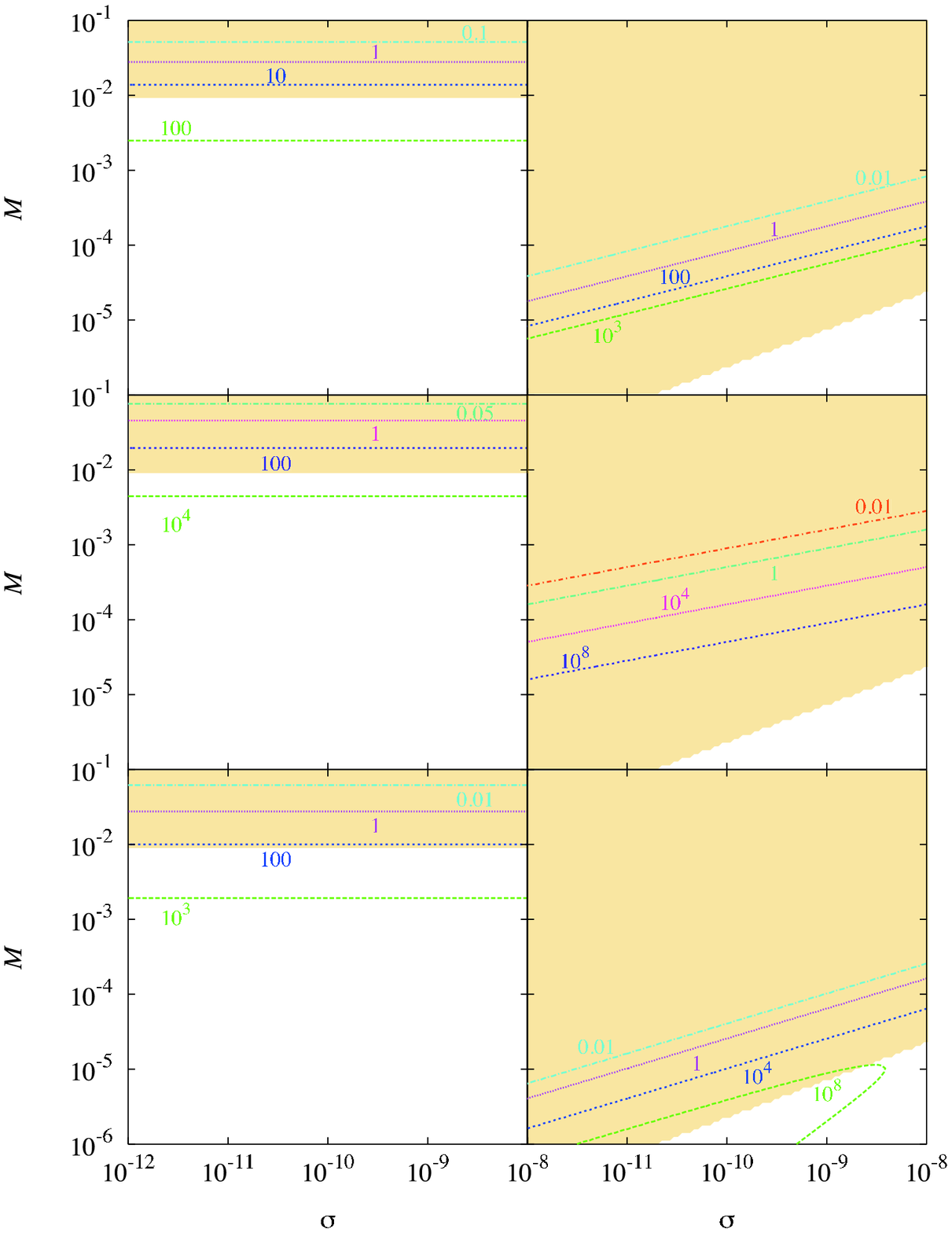}}
\end{tabular}
\caption{Contours of $f_{\rm NL}$ (top panels), $\tau_{\rm NL}$
  (middle panels) and $g_{\rm NL}$ (bottom panels) in the
  $\sigma$--$M$ plane for the chaotic inflation model with quadratic
  and quartic potential.  Left (right) panels are for case A
  (B). Shaded region is excluded by WMAP5.}
\label{chao24nl}
\end{center}
\end{figure}

Next we consider chaotic inflation with the quartic potential.  As a
simple possibility, one usually assumes that the quartic term alone
exists in the potential for the inflaton.  Since the energy density of
inflaton oscillations in such a potential decreases in the same way as
that of radiation, no fluctuation is generated via modulated reheating
scenario as discussed previously.  However, there are some interesting
models \cite{Kadota:2007nc,Kadota:2008pm} where the potential during
inflation is given by the quartic form, but the quadratic term becomes
effective when the inflaton starts to oscillate. In fact, this feature
comes from the curved trajectory in the multiple fields
configuration. But it can be well characterized by considering a
single field with the potential given by the sum of the quadratic term
and the quartic one: $V(\phi)=\frac{m^2}{2}
\phi^2+\frac{\lambda}{4}\phi^4$.  Above situation can be realized by
taking $m= \sqrt{\lambda}M_{\rm pl}$ with $\lambda$ being fixed by
WMAP normalization.  Hence we adopt this relation in the following
analysis for definiteness.  In this case, quartic term dominates
during inflation ($\phi \gtrsim M_{\rm pl}$), but the quadratic one
dominates after inflation.

This model is also interesting in another aspect.  In the absence of
the quadratic term, we have $Q=0$ and the curvature perturbations are
not generated by the modulation of the decay rate.  Then the curvature
perturbations originate solely from the inflaton fluctuations even
when the decay rate fluctuates, which is completely ruled out from
WMAP 5-year data because of the too large tensor-to-scalar ratio.
However, if we include the quadratic one, the curvature perturbations
can be additionally generated when inflaton decays.  If these
perturbations dominate the total curvature perturbations, the quartic
inflation model may still satisfy the observational constraints, with
large amount of the non-Gaussianity which could be tested by the
future observations.

Fig.~\ref{chao24nsr} show contour plots of $n_s$ and $r$ in this case,
respectively.  Shaded regions are excluded by the WMAP 5-year
constraints on $(n_s,r)$.  We see that when the modulus contributions
become dominant, $n_s$ gets closer to unity and $r$ becomes very
small.  As a result, $(n_s,r)$ enters the allowed region of WMAP
5-year data.  The slopes of the contours are essentially the same as
in the previous case where $V(\phi)=\frac{m^2}{2}\phi^2$.

In Fig.~\ref{chao24nl}, we show the contour plots of the three
non-linearity parameters.  As for the case A, there is a large
parameter space where $(n_s,r)$ is in the allowed region of WMAP
5-year data while generating large amount of non-Gaussianity $f_{\rm
  NL} ={\cal O}(10\sim 100)$ which could be tested by the upcoming
observations.

\begin{figure}[!h]
%\vspace{-1.2cm}
\begin{center}
\begin{tabular}{cc}
\resizebox{150mm}{!}{\includegraphics{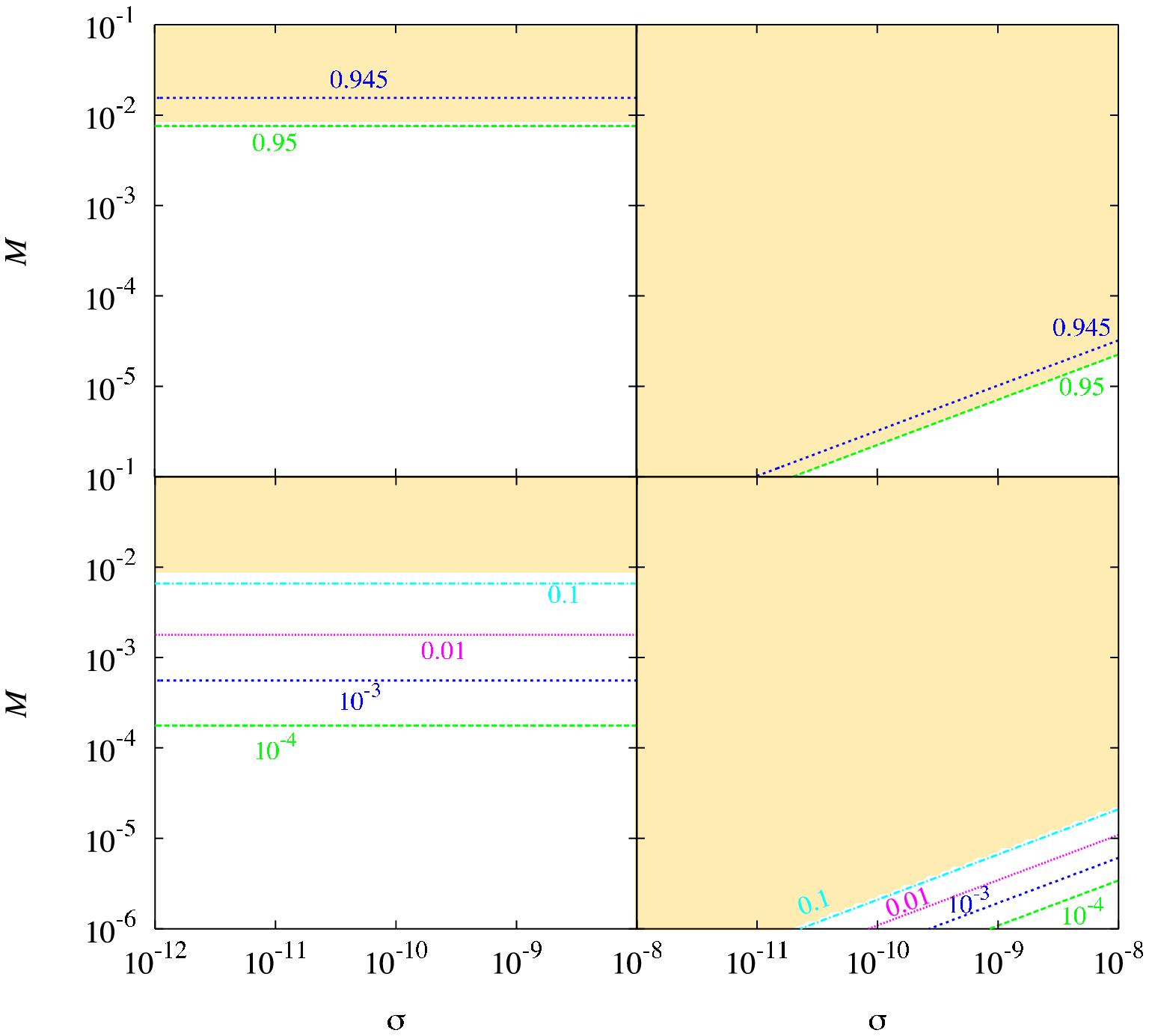}}
\end{tabular}
\caption{Contours of $n_s$ (top panels) and $r$ (bottom panels) in the
  $\sigma$--$M$ plane for the chaotic inflation model with sextic
  potential.  Left (right) panels are for case A (B). Shaded region is
  excluded by WMAP5.}
\label{chao6nsr}
\end{center}
\end{figure}

%%%%%%%%%%%%%%%%%%%%%%%%%%%%%%%%%%%%

\subsection{Chaotic inflation : $V(\phi)=\frac{V_0}{6} {\left( \frac{\phi}{M_{\rm pl}} \right)}^6$}

Let us next consider chaotic inflation model whose potential is given
by
\begin{eqnarray}
V(\phi)=\frac{V_0}{6} {\left( \frac{\phi}{M_{\rm pl}} \right)}^6.
\end{eqnarray}
This model is also completely ruled out by WMAP 5-year data if the
curvature perturbations solely originate from the inflaton
fluctuations.

Fig.~\ref{chao6nsr} show the contour plots of $n_s$ and $r$ in this
case, respectively.  We see that as in the case of the quadratic and
quartic potential, $(n_s,r)$ enters the allowed region of WMAP 5-year
data due to the slight shift of $n_s$ to unity and the significant
suppression of $r$ when the modulus contributions become dominant.

\begin{figure}[!h]
\vspace{-1.2cm}
\begin{center}
\begin{tabular}{cc}
\resizebox{150mm}{!}{\includegraphics{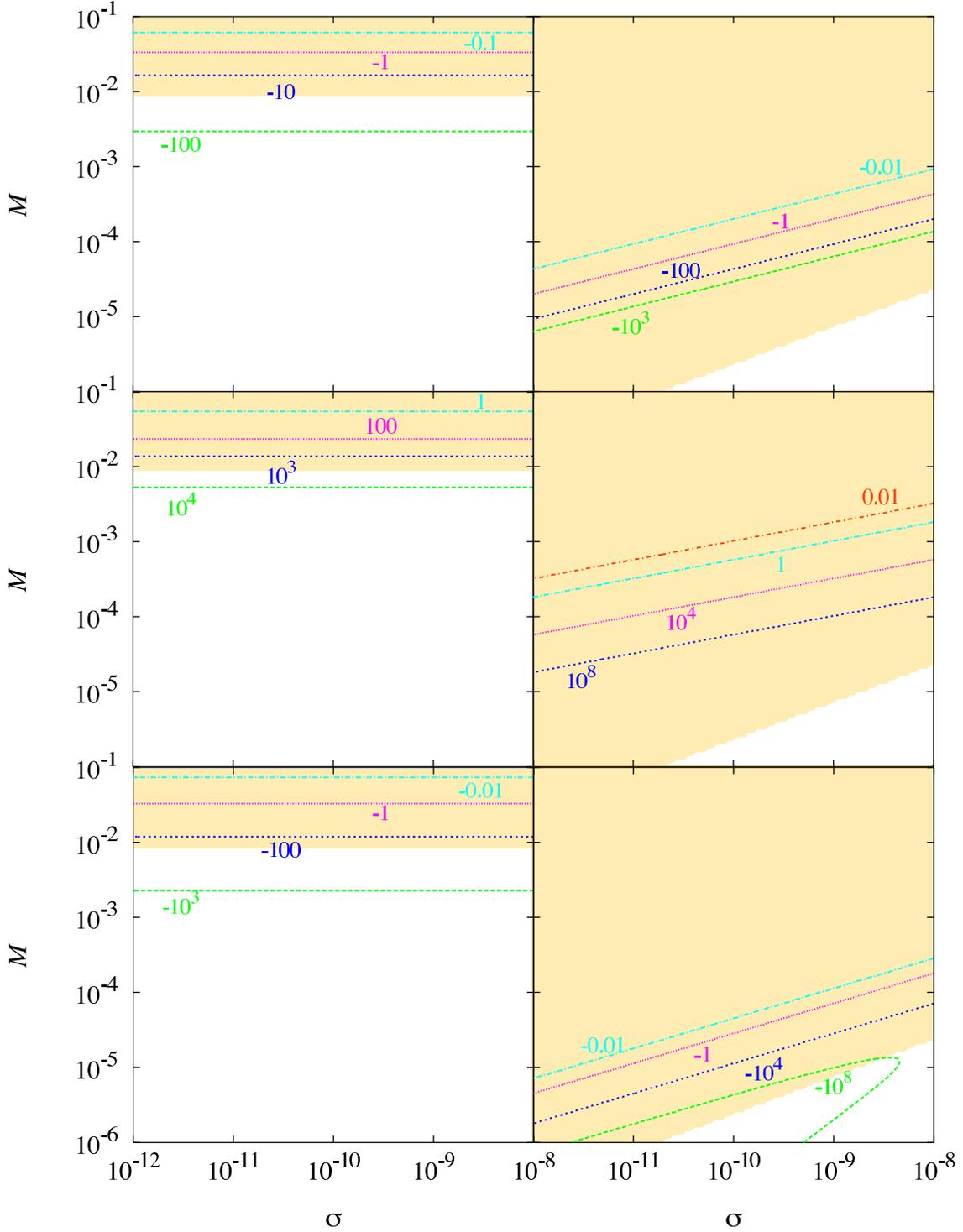}}
\end{tabular}
\caption{Contours of $f_{\rm NL}$ (top panels), $\tau_{\rm NL}$
  (middle panels) and $g_{\rm NL}$ (bottom panels) in the
  $\sigma$--$M$ plane for the chaotic inflation model with sextic
  potential.  Left (right) panels are for case A (B). Shaded region is
  excluded by WMAP5.}
\label{chao6nl}
\end{center}
\end{figure}

In Fig.~\ref{chao6nl}, we show the contour plots of the three
non-linearity parameters.  For the sextic potential, $f_{\rm NL}$
becomes negatively large when the modulus contributions are dominant,
as discussed in the previous section.  If we take $\beta <0$ instead
of $\beta >0$, then we have positively large $f_{\rm NL}$ just like
the case in the quadratic and quartic potential.

%%%%%%%%%%%%%%%%%%%%%%%%%%%
\section{Summary}\label{sec:summary}
%%%%%%%%%%%%%%%%%%%%%%%%%%%

We have investigated a mixed scenario where both fluctuations from the
inflaton and modulated reheating are responsible for cosmic density
fluctuations today.  First we summarized the decay rate of the
inflaton for various interactions and inflaton potentials, whose
details are presented in Appendix \ref{sec:decay}.  Then we gave
expressions for the spectral index, tensor-to-scalar ratio and
non-linearity parameters such as $f_{\rm NL},\tau_{\rm NL}$ and
$g_{\rm NL}$.  We found that while the non-linearity between $\zeta$
and $\delta \Gamma$ generates $f_{\rm NL}={\cal O}(1\sim 10)$, the
non-linearity between $\Gamma$ and the modulus can generate large
$f_{\rm NL} \ge {\cal O}(100)$.  We also derived the consistency
relation Eq.~\eqref{eq:consistency} among the non-linearity parameters
which is applicable for the cases where the decay rate for the
inflaton is given as Eq.~\eqref{eq:decay1}.  If the potential around
the minimum is quadratic, then $g_{\rm NL}$ takes the same sign as
$f_{\rm NL}$.  This is in sharp contrast to the situation where the
mixed model of the inflaton and the curvaton yields a negative $g_{\rm
  NL}$ when $f_{\rm NL}$ is large positive.

We have also studied the inflationary parameters including
non-linearity parameters assuming some inflation models.  Since the
new inflation model usually gives a extremely small value of
$\epsilon$, we have $R \ll 1$.  Thus, the addition of fluctuations
from modulated reheating does not significantly change the prediction
of the property of fluctuations in the case of new
inflation. Therefore, in this paper, we have considered chaotic
inflation models with several polynomials and some types of decay rate
for an illustrational purpose. In the case with the chaotic inflation
model, the contribution from fluctuations of modulated reheating makes
non-Gaussianity larger, the spectral index closer to scale-invariant
and tensor-to-scalar ratio more suppressed.  This helps to liberate
models of inflation such as the sextic potential one because this
model has been excluded by the data because of its too red-tilted
spectrum and too large tensor-to-scalar ratio.  The chaotic inflation
with quartic potential has also in fact been excluded by the
data. Since fluctuations are not generated by modulated reheating in a
simple quartic potential case, this model cannot be made viable by
just adding a contribution from the modulated reheating.  However,
even if the inflaton is driven by a quartic term, another term such as
a quadratic one can become effective during reheating stage after
inflation. This kind of situation can be realized by assuming $V(\phi)
= (\lambda / 4) \phi^4 + (1/2) m^2 \phi^2$ with $m = \sqrt{\lambda}
M_{\rm pl}$, which was discussed in this paper. In such a case,
fluctuations from modulated reheating can affect the total
fluctuations and liberate the model. In addition, non-Gaussianity can
also be large in this case too.  We have also investigated the case
with the quadratic inflation model and showed that non-Gaussianity can
be large in this model by just adding some contribution from modulated
reheating without conflicting the observations of the spectral index
and tensor modes.

In most works so far, observational consequences for various
generation mechanisms of primordial fluctuations have been discussed
in each separately.  However, different mechanisms can be in general
responsible for today's cosmic density fluctuations simultaneously so
it is of great importance to investigate a mixed model such as the one
we discussed in this paper.  Precise cosmological observations
expected in the near future to hunt for primordial non-Gaussianity may
reveal such interesting scenarios in the early universe.

\bigskip
\bigskip

\noindent {\bf Acknowledgments:} 
T.S. is grateful to Masahiro Kawasaki and Fuminobu Takahashi for helpful discussion.
T.S. also thanks to the computer system at the Yukawa Institute for
Theoretical Physics, Kyoto University, for the numerical calculations.
This work is supported in part by the JSPS Postdoctoral Fellowship for
Research Abroad (K.I.), the Sumitomo Foundation (T.T.), and the
Grant-in-Aid for Scientific Research from the Ministry of Education,
Science, Sports, and Culture of Japan No.\,19740145 (T.T.),
No.\,18740157, and No.\,19340054 (M.Y.).

\pagebreak

\appendix 

\noindent
{\bf \Large Appendix}

%%%%%%%%%%%%%%%%%%%%%%%%%%%%%%%%%
\section{Calculation of the decay rate}\label{sec:decay}
%%%%%%%%%%%%%%%%%%%%%%%%%%%%%%%%%

In this appendix, we calculate the decay rate of the inflaton to
lighter particles for three different types of interactions
\cite{Linde:2005ht,Shtanov:1994ce,Greene:1997fu}.  We also consider
three different types of potentials around the minimum: quadratic,
quartic and sextic potentials.  In the following, we approximate the
background space-time as Minkowski space because the energies of
created particles from the inflaton are much higher than the Hubble
parameter.  In this appendix, we follow the notations of
Ref.~\cite{Wein}.

\subsection{Yukawa interaction: ${\cal L}_{\rm int}=-y \phi {\bar \psi}\psi$}

We regard $\phi$ as the classical field which is spatially homogeneous
but oscillates in time $t$.  Then the interaction part of the
Hamiltonian is given by
\begin{eqnarray}
{\hat V}=y\phi (t) \int d^3 x~{\bar \psi}\psi.
\end{eqnarray}
$\psi$ in the interaction picture is equal to that in the Heisenberg
picture when the interaction is absent.  Hence we write $\psi$ as
\begin{eqnarray}
\psi (x)=\sum_\sigma \int \frac{d^3 p}{ {(2\pi)}^{3/2}} \left( u({\vec p},\sigma) e^{ipx} b_{ {\vec p},\sigma}+v({\vec p},\sigma) e^{-ipx} b^{c\dagger}_{{\vec p},\sigma} \right),
\end{eqnarray}
where $\sigma$ denotes the helicity and 
$b_{ {\vec p},\sigma}$ and $b^c_{{\vec p},\sigma}$ satisfy the following relations,
\begin{eqnarray}
\{ b_{{\vec p},\sigma},~b_{{\vec q},\sigma'}^\dagger \}=\{ b^c_{{\vec p},\sigma},~b_{{\vec q},\sigma'}^{c \dagger} \}=\delta ({\vec p}-{\vec q}) \delta_{\sigma \sigma'},~~~~~\{ b_{{\vec p},\sigma},~b_{{\vec q},\sigma'} \}=\{ b^c_{{\vec p},\sigma},~b^c_{{\vec q},\sigma'} \}=0.
\end{eqnarray}
Also, $u({\vec p},\sigma)$ and $v({\vec p},\sigma)$ satisfy the
following equations,
\begin{eqnarray}
(ip^\mu \gamma_\mu+m_\psi) u({\vec p},\sigma)=0,~~~~~(-ip^\mu \gamma_\mu+m_\psi) v({\vec p},\sigma)=0.
\end{eqnarray}

Let us write the period of the oscillations of $\phi(t)$ as $T$ and
Fourier expand $\phi(t)$ as
\begin{eqnarray}
\phi (t)=\sum_{n=-\infty}^\infty~\phi_n e^{-i \omega n t},
\end{eqnarray}
where $\omega \equiv 2\pi /T$.

Using these, the transition amplitude from the initial state
$|i\rangle =| 0 \rangle$ at $t=-\infty$ to the final two-particle
state $|f\rangle =b_{\vec p, \sigma}^\dagger b_{\vec q,
  \sigma'}^{c\dagger} |0 \rangle$ at $t=\infty$ is given by
\begin{eqnarray}
-i\int_{-\infty}^\infty dt~\langle f| V(t) |i \rangle=-2\pi iy \delta ({\vec p}+{\vec q}) ~{\bar u}({\vec p},\sigma) v( -{\vec p},\sigma') \sum_{n=-\infty}^\infty \phi_n \delta (2E_p-n\omega),
\end{eqnarray}
where $E_p \equiv \sqrt{p^2+m_\psi^2}$ is the energy of the $\psi$
particle.  Then $\Gamma$, the transition rate per unit time and unit
volume, is given by
\begin{eqnarray}
\Gamma&=&y^2 \sum_{n=1}^\infty {|\phi_n|}^2 \int \frac{d^3 p}{ {(2\pi)}^2} \delta (2E_p-n\omega) \sum_{\sigma \sigma'} {| {\bar u} ({\vec p},\sigma) v(-{\vec p},\sigma')|}^2 \nonumber \\
&=&\frac{y^2}{4\pi} \omega^2 \sum_{n=1}^\infty n^2 {|\phi_n|}^2 \nonumber \\
&=&\frac{y^2}{8\pi} \langle {\dot \phi}^2 \rangle.
\end{eqnarray}
Here $\langle \cdots \rangle$ denotes the average of $\cdots$ over one
period of the oscillations.  From the first line to the second one, we
have used the relation,
\begin{eqnarray}
\sum_{\sigma \sigma'} {| {\bar u} ({\vec p},\sigma) v(-{\vec p},\sigma')|}^2=\frac{2p^2}{E_p^2},
\end{eqnarray}
and assumed $\psi$ is massless, i.e., $m_\psi=0$.

$\Gamma$ is the production rate of two $\psi$-particle from the
vacuum, not the decay rate of $\phi$-field.  The decay rate of the
$\phi$ energy, which we denote as $\Gamma_\phi$, can be obtained from
the energy conservation,
\begin{eqnarray}
\rho_\phi \Gamma_\phi \Delta t= E \Gamma \Delta t.
\end{eqnarray}
The left-hand side denotes the energy loss of the $\phi$-field during
the infinitesimal time $\Delta t$ while the right-hand side the
energy gain of two $\psi$-particles.  $E$ is the expectation value of
the energy of the final two-particle state.  From this equation,
$\Gamma_\phi$ can be written as
\begin{eqnarray}
\Gamma_\phi= \frac{y^2}{8\pi} E \frac{ \langle {\dot \phi}^2 \rangle }{\rho_\phi}.
\end{eqnarray}

From the definition of $E$, it can be written as
\begin{eqnarray}
E=\frac{ \sum_{n, \sigma, \sigma'} \int d^3 p d^3 q \delta ({\vec p}+{\vec q}) E_f \delta (E_f-n\omega) {| {\cal M}_n |}^2 }{\sum_{n, \sigma, \sigma'} \int d^3 p d^3 q \delta ({\vec p}+{\vec q}) \delta (E_f-n\omega) {| {\cal M}_n |}^2},
\end{eqnarray}
where $E_f$ is the energy of the final state and ${\cal M}_n$ is defined
by
\begin{eqnarray}
{\cal M}_n = -2\pi y i {\bar u} ({\vec p},\sigma) v(-{\vec p},\sigma') \phi_n.
\end{eqnarray}
Substituting this into the equation above yields
\begin{eqnarray}
E=\frac{ \sum_{n=1}^\infty {| \phi_n |}^2 {(n\omega )}^3}{\sum_{n=1}^\infty {| \phi_n |}^2 {(n\omega )}^2}.
\end{eqnarray}

Let us define the numerical factor $\alpha$ by
\begin{eqnarray}
\alpha=\frac{ \sum_{n=1}^\infty {| \phi_n |}^2 n^3}{\sum_{n=1}^\infty {| \phi_n |}^2 n^2}.
\end{eqnarray}
Then $\Gamma_\phi$ can be written as
\begin{eqnarray}
\Gamma_\phi=\frac{y^2}{8\pi} \omega \alpha \frac{ \langle {\dot \phi}^2 \rangle }{\rho_\phi}. \label{gamma}
\end{eqnarray}
Hence once we specify the motion of $\phi (t)$, we can immediately
calculate $\Gamma_\phi$ using Eq.~(\ref{gamma}).

\subsubsection{Quadratic potential:$V(\phi)=\frac{m^2}{2}\phi^2$}

In this case, $\phi(t)$ can be written as
\begin{eqnarray}
\phi (t)=\phi_0 \cos (m t), \label{qua1}
\end{eqnarray}
where the frequency, $m$, is just equal to the mass of the inflaton.

Then, we have
\begin{eqnarray}
\alpha=1,~~~~~\frac{ \langle {\dot \phi}^2 \rangle }{\rho_\phi}=1.
\end{eqnarray}
Hence $\Gamma_\phi$ is given by
\begin{eqnarray}
\Gamma_\phi=\frac{y^2}{8\pi} m.
\end{eqnarray}

\subsubsection{Quartic potential:$V(\phi)=\frac{\lambda}{4}\phi^4$}

In this case,
$\phi(t)$ can be written as
\begin{eqnarray}
\phi(t)=\frac{\sqrt{\pi} \Gamma(\frac{3}{4})}{\Gamma (\frac{5}{4} )}\phi_0 \sum_{n=1}^\infty \left( e^{i(2n-1)\omega t}+e^{-i(2n-1)\omega t} \right) \frac{e^{-\frac{\pi}{2} (2n-1)}}{1+e^{-\pi (2n-1)}}, \label{qua2}
\end{eqnarray}
where the frequency $\omega$ is given by
\begin{eqnarray}
\omega =\frac{1}{2} \sqrt{\frac{\pi}{6}} \frac{\Gamma(\frac{3}{4})}{\Gamma(\frac{5}{4})} m_\phi^{\rm eff}.
\end{eqnarray}
$m_\phi^{\rm eff}$ is the effective mass of $\phi$ defined by
\begin{eqnarray}
m_\phi^{\rm eff} \equiv \sqrt{3\lambda} \phi_0.
\end{eqnarray}

Using these,
we find
\begin{eqnarray}
\alpha \approx 1.036,~~~~~\frac{ \langle {\dot \phi}^2 \rangle }{\rho_\phi}=\frac{4}{3}.
\end{eqnarray}

Then the decay rate of the inflaton can be written as
\begin{eqnarray}
\Gamma_\phi=A_2 \frac{y^2}{8\pi}m_\phi^{\rm eff},
\end{eqnarray}
where $A_2 \approx 0.676$ is a numerical constant.

\subsubsection{Sextic potential:$V(\phi)=\frac{V_0}{6M_{\rm pl}^6} \phi^6$}

In this case, 
the frequency of the inflaton oscillations is given by
\begin{eqnarray}
\omega =\frac{1}{2} \sqrt{\frac{\pi}{15}} \frac{\Gamma(\frac{2}{3})}{\Gamma(\frac{7}{6})} m_\phi^{\rm eff}.
\end{eqnarray}
$m_\phi^{\rm eff}$ is the effective mass of $\phi$ defined by
\begin{eqnarray}
m_\phi^{\rm eff} \equiv \frac{\sqrt{5 V_0}}{M_{\rm pl}^3} \phi_0^2.
\end{eqnarray}

To determine $\alpha$,
we numerically solved the equation of motion for $\phi$ over
one period of the oscillations.
The result is
\begin{eqnarray}
\alpha \approx 1.0897,~~~~~\frac{ \langle {\dot \phi}^2 \rangle }{\rho_\phi}=\frac{3}{2}.
\end{eqnarray}
Then the decay rate of the inflaton can be written as
\begin{eqnarray}
\Gamma_\phi=A_3 \frac{y^2}{8\pi}m_\phi^{\rm eff},
\end{eqnarray}
where $A_3 \approx 0.546$ is a numerical constant.

\subsection{Interaction with scalar field $\chi$: ${\cal L}_{\rm int}=-M \phi \chi \chi$}

In this case, the interaction part of the Hamiltonian is given by
\begin{eqnarray}
{\hat V}=M \phi(t) \int d^3x ~{\hat \chi} {\hat \chi}.
\end{eqnarray}
Let us expand $\chi$ as
\begin{eqnarray}
\chi (x)=\int \frac{d^3p}{ {(2\pi)}^{3/2} \sqrt{2E_p}} \left( e^{ipx} a_{\vec p}+e^{-ipx} a_{\vec p}^\dagger \right).
\end{eqnarray}

Then the transition amplitude from the initial state $|i\rangle =
|0\rangle$ at $t=-\infty$ to the final two-particle state
$|f\rangle=a_{\vec p}^\dagger a_{\vec q}^\dagger |0\rangle$ is given by
\begin{eqnarray}
-i \int_{-\infty}^\infty dt~\langle f| V(t) |i\rangle=-2\pi i M ~\delta ({\vec p}+{\vec q})\sum_{n=-\infty}^\infty \frac{\phi_n}{E_p} \delta (2E_p-n\omega).
\end{eqnarray}
The transition rate per unit time and unit volume becomes
\begin{eqnarray}
\Gamma&=&\frac{M^2}{4\pi} \sum_{n=1}^\infty {|\phi_n|}^2 \nonumber \\
&=&\frac{M^2}{8\pi} \langle \phi^2 \rangle.
\end{eqnarray}
The corresponding decay rate of the inflaton is given by
\begin{eqnarray}
\Gamma_\phi=\frac{M^2}{8\pi} E \frac{\langle \phi^2 \rangle}{\rho_\phi},
\end{eqnarray}
where $E$, the mean energy of the two-particle state, is given by
\begin{eqnarray}
E=\frac{ \sum_{n=1}^\infty {n\omega | \phi_n |}^2 }{\sum_{n=1}^\infty {| \phi_n |}^2 }.
\end{eqnarray}

Introducing the dimensionless number $\beta$ by
\begin{eqnarray}
\beta \equiv \frac{ \sum_{n=1}^\infty {n| \phi_n |}^2 }{\sum_{n=1}^\infty {| \phi_n |}^2 },
\end{eqnarray}
$\Gamma_\phi$ becomes
\begin{eqnarray}
\Gamma_\phi=\frac{M^2}{8\pi} \beta \omega \frac{\langle \phi^2 \rangle}{\rho_\phi}.
\end{eqnarray}

\subsubsection{Quadratic potential:$V(\phi)=\frac{m^2}{2}\phi^2$}

Using Eq.~(\ref{qua1}), we have
\begin{eqnarray}
\beta=1,~~~~~\frac{ \langle \phi^2 \rangle }{\rho_\phi}=\frac{1}{m^2}.
\end{eqnarray}
Hence $\Gamma_\phi$ is given by
\begin{eqnarray}
\Gamma_\phi=\frac{M^2}{8\pi m}.
\end{eqnarray}

\subsubsection{Quartic potential:$V(\phi)=\frac{\lambda}{4}\phi^4$}

Using Eq.~(\ref{qua2}), we find
\begin{eqnarray}
\beta \approx 1.004,~~~~~\frac{ \langle \phi^2 \rangle }{\rho_\phi}\approx \frac{5.48}{{(m_\phi^{\rm eff})}^2}.
\end{eqnarray}
Then the decay rate of the inflaton can be written as
\begin{eqnarray}
\Gamma_\phi=B_2\frac{M^2}{8\pi m_\phi^{\rm eff}},
\end{eqnarray}
where $B_2 \approx 2.693$ is a numerical constant.

\subsubsection{Sextic potential:$V(\phi)=\frac{V_0}{6M_{\rm pl}^6} \phi^6$}

In this case, we numerically found
\begin{eqnarray}
\beta \approx 1.010,~~~~~\frac{ \langle \phi^2 \rangle }{\rho_\phi}\approx \frac{12.93}{{(m_\phi^{\rm eff})}^2}.
\end{eqnarray}
Then the decay rate of the inflaton can be written as
\begin{eqnarray}
\Gamma_\phi=B_3 \frac{M^2}{8\pi m_\phi^{\rm eff}},
\end{eqnarray}
where $B_3 \approx 4.362$ is a numerical constant.

\subsection{Interaction with scalar field $\chi$: ${\cal L}_{\rm int}=-h \phi^2 \chi^2$}

In this case,
the interaction part of the Hamiltonian is given by
\begin{eqnarray}
{\hat V}=h \phi^2(t) \int d^3x ~{\hat \chi} {\hat \chi}.
\end{eqnarray}

Let us define $\zeta (t)$ as
\begin{eqnarray}
\phi^2 (t)-\langle \phi^2 \rangle=\sum_{n=-\infty}^\infty \zeta_n e^{-in \omega t}.
\end{eqnarray}
Note that $\omega$ is the frequency of $\zeta(t)$.  Then the transition
rate per unit time and unit volume is
\begin{eqnarray}
\Gamma&=&\frac{h^2}{4\pi} \sum_{n=1}^\infty {|\zeta_n|}^2 \nonumber \\
&=&\frac{h^2}{8\pi} \left( \langle \phi^4 \rangle- \langle \phi^2 \rangle^2 \right).
\end{eqnarray}
Introducing the dimensionless quantity $\gamma$ as
\begin{eqnarray}
\gamma \equiv \frac{ \sum_{n=1}^\infty {n| \zeta_n |}^2 }{\sum_{n=1}^\infty {| \zeta_n |}^2 },
\end{eqnarray}
the inflaton decay rate is given by
\begin{eqnarray}
\Gamma_\phi=\frac{h^2}{8\pi} \gamma \omega \frac{\langle \phi^4 \rangle- \langle \phi^2 \rangle^2}{\rho_\phi}.
\end{eqnarray}

\subsubsection{Quadratic potential:$V(\phi)=\frac{m^2}{2}\phi^2$}

Using Eq.~(\ref{qua1}), we have
\begin{eqnarray}
\gamma=1,~~~~~\frac{ \langle \phi^4 \rangle-\langle \phi^2 \rangle^2 }{\rho_\phi}=\frac{\phi_0^2}{4m^2}.
\end{eqnarray}
Hence $\Gamma_\phi$ is given by
\begin{eqnarray}
\Gamma_\phi=\frac{h^2 \phi_0^2}{32\pi m}=\frac{h^2}{16\pi m^3} \rho_\phi.
\end{eqnarray}

\subsubsection{Quartic potential:$V(\phi)=\frac{\lambda}{4}\phi^4$}

Using Eq.~(\ref{qua2}), we find
\begin{eqnarray}
\gamma \approx 1.007,~~~~~\frac{ \langle \phi^4 \rangle-\langle \phi^2 \rangle^2 }{\rho_\phi}\approx \frac{0.50}{\lambda}.
\end{eqnarray}
Then the decay rate of the inflaton can be written as
\begin{eqnarray}
\Gamma_\phi=C_2 \frac{h^2}{8\pi {( m_\phi^{\rm eff })}^3} \rho_\phi,
\end{eqnarray}
where $C_2 \approx 8.86$ is a numerical constant.

\subsubsection{Sextic potential:$V(\phi)=\frac{V_0}{6M_{\rm pl}^6} \phi^6$}

In this case, we numerically found
\begin{eqnarray}
\gamma \approx 1.019,~~~~~\frac{ \langle \phi^4 \rangle-\langle \phi^2 \rangle^2 }{\rho_\phi}\approx \frac{0.73}{V_0 \phi_0^2}.
\end{eqnarray}
Then the decay rate of the inflaton can be written as
\begin{eqnarray}
\Gamma_\phi=C_3 \frac{h^2}{8\pi {( m_\phi^{\rm eff })}^3} \rho_\phi,
\end{eqnarray}
where $C_3 \approx 37.26$  is a numerical constant.

%%%%%%%%%%%%%%%%%%%%%%%%%%%%%
\section{Calculation of $Q(x)$}\label{sec:app_Q}
%%%%%%%%%%%%%%%%%%%%%%%%%%%%%

In this section, we will provide explicit form of $Q(x)$ for three types
of inflaton potential, quadratic $V \propto \phi^2$, quartic $V
\propto \phi^4$ and sextic $V \propto \phi^6$, and for three types
of interactions between the inflaton and matter particles, Yukawa interactions,
three-point interactions ${\cal L}_{\rm int}=-M\phi \chi^2$ and
four-point interactions ${\cal L}_{\rm int}=-h \phi^2 \chi^2$.

\subsection{Quadratic potential}

If the inflaton potential is quadratic,
i.e. $V(\phi)=\frac{m^2}{2}\phi^2$, then we have $m_{\rm eff}=m$.
Hence the decay rate becomes independent of time for Yukawa
interactions and three-point interactions ${\cal L}_{\rm int}=-M\phi
\chi^2$.  In this case, it was shown in Ref.~\cite{Suyama:2007bg} that
$Q(x)=-\frac{1}{6} \log x$.  Meanwhile, if the dominant decay occurs
through the four-point interactions, then the inflaton decay rate
decreases as $\Gamma_\phi \propto \rho_\phi$.  From the Friedmann
equation $H^2 \propto \rho_\phi$, the ratio $\Gamma_\phi/H$ decreases
in proportional to $a^{-3/2}$.  Since $\Gamma_\phi$ is smaller than
$H$ at initial time, that is, at the end of inflation, $\Gamma_\phi$
never becomes larger than $H$.  Hence the universe is never reheated
only by the four-point interactions, which cannot realize the hot big
bang cosmology. Thus we do not consider this case.

\subsection{Quartic potential}

If the inflaton potential is quartic, $\rho_\phi$ decreases in
proportional to $a^{-4}$.  Hence the universe expands in the same way
as the radiation dominated universe.  From the definition of $Q(x)$
(see Eq.~(\ref{e1})), we find
\begin{eqnarray}
Q(x)=0,
\end{eqnarray}
for all three types of interactions.  Note that unlike in the case of
the quadratic potential, the universe can be reheated only by the
four-point interactions in this case.

\subsection{Sextic potential}

If the inflaton potential is sextic, $\rho_\phi$ decays in
proportional to $e^{-9N/2}$.  Then the decay rate of the inflaton
evolves as $\propto e^{-3N/2},~e^{3N/2}$ and constant for Yukawa
interactions, three-point interactions ${\cal L}_{\rm int}=-M\phi
\chi^2$ and four-point interactions ${\cal L}_{\rm int}=-h \phi^2
\chi^2$, respectively.  Since the Hubble parameter decays faster
than $e^{-2N}$, the ratio $\Gamma_\phi/H$ grows in time for any
interactions.  Hence the universe eventually becomes the radiation
dominated universe.

Since the most of radiation is produced when $H=\Gamma_\phi$, which we
checked by a numerical calculation, in what follows, we will use the
so-called sudden decay approximation, where the inflaton decays
instantaneously when the decay rate becomes equal to the Hubble
parameter.  Under this approximation, the background equations can be
written as
\begin{eqnarray}
&&\frac{d\rho_r}{dN}+4\rho_r=\frac{\Gamma_\phi}{H} \rho_\phi \delta (N-N_d), \label{sexticrad1}\\
&&\frac{d\rho_\phi}{dN}+\frac{9}{2} \rho_\phi=-\frac{\Gamma_\phi}{H} \rho_\phi \delta (N-N_d), \\
&&H^2=\frac{1}{3M_{\rm pl}^2} (\rho_\phi+\rho_r).
\end{eqnarray}
Integrating Eq.~(\ref{sexticrad1}) from $N=0$ which corresponds to
$t=t_c$ to $N=N_f>N_d$, we find
\begin{eqnarray}
N(t_f,t_c)=\frac{1}{4} \log \frac{\rho_c}{\rho_f}+\frac{1}{2} \log \frac{\Gamma_\phi (t_c)}{H_c}+N_d.
\end{eqnarray}
Comparing this equation with Eq.~(\ref{e1}), $Q(x)$ can be written as
\begin{eqnarray}
Q=\frac{1}{2} \log \frac{\Gamma_\phi (N_d)}{H_c}+N_d. \label{sexticQ1}
\end{eqnarray}

\subsubsection{Yukawa interactions}

In this case, the decay rate can be written as
$\Gamma_\phi=\Gamma_\phi(t_c) e^{-3N/2}$.  Hence from the equation
$H=\Gamma_\phi$, we find that $N_d$ is given by
\begin{eqnarray}
N_d =-\frac{4}{3} \log \frac{\Gamma_\phi (t_c)}{H_c}.
\end{eqnarray}
Substituting this into Eq.~(\ref{sexticQ1}) gives
\begin{eqnarray}
Q=\frac{1}{6} \log \frac{\Gamma_\phi (t_c)}{H_c}.
\end{eqnarray}

\subsubsection{Three-point interactions ${\cal L}_{\rm int}=-M\phi \chi^2$}

In this case, the decay rate is $\Gamma_\phi=\Gamma_\phi (t_c)
e^{3N/2}$.  Then the corresponding $Q$ is given by
\begin{eqnarray}
Q=\frac{1}{30} \log \frac{\Gamma_\phi (t_c)}{H_c}.
\end{eqnarray}

\subsubsection{Four-point interactions ${\cal L}_{\rm int}=-h\phi^2 \chi^2$}

In this case, the decay rate is $\Gamma_\phi=\Gamma_\phi (t_c)$.  Then
the corresponding $Q$ is given by
\begin{eqnarray}
Q=\frac{1}{18} \log \frac{\Gamma_\phi (t_c)}{H_c}.
\end{eqnarray}

\end{document}